\documentclass[12pt]{article}

\usepackage{jheppub} 
\usepackage{float}
\usepackage{dsfont}
\usepackage{amsmath,bbm,blkarray,bm}
\usepackage[vcentermath]{youngtab}
\usepackage{slashed}
\usepackage{bbold}
\usepackage{booktabs}
\usepackage{multirow}
\usepackage{tikz}
\usepackage{changepage}
\usepackage{makecell}
\usepackage{tikz-feynman}
\usepackage{subcaption}
\usepackage{color}
\usepackage[numbers,sort&compress]{natbib}
\usepackage[T1]{fontenc} 

\usepackage{adjustbox}
\usepackage{cancel}

\newcommand{\1}{\mathbb{1}}

\newcommand{\be} {\begin{equation}}
\newcommand{\ee} {\end{equation}}
\newcommand{\bea} {\begin{eqnarray}}
\newcommand{\eea} {\end{eqnarray}}

\newcommand{\eval}[1]{\big<#1\big>}

\newcommand{\T}[1]{\tilde{#1}}

\allowdisplaybreaks

\title{Renormalization scheme factorization of one-loop Fierz identities}

\author[a]{Jason~Aebischer,}
\author[a]{Marko Pesut,}
\author[a]{Zachary Polonsky}

\affiliation[a]{Physik-Institut, Universit\"at Z\"urich, CH-8057 Z\"urich, Switzerland}

\emailAdd{jason.aebischer@physik.uzh.ch}
\emailAdd{marko.pesut@physik.uzh.ch}
\emailAdd{zach.polonsky@physik.uzh.ch}

\abstract{
	We present a proof of the factorization of renormalization scheme in one-loop-corrected Fierz identities. This scheme factorization facilitates the simultaneous transformation of operator basis and renormalization scheme using only relations between physical operators; the evanescent operators in the respective bases may be chosen entirely independently of each other. The relations between evanescent operators in the two bases is automatically accounted for by the corrected Fierz identities. We illustrate the utility of this result with a two-loop anomalous dimension matrix computation using the Naive-Dimensional Regularization scheme, which is then transformed via one-loop Fierz identities to the known result in the literature given in a different basis and calculated in the Larin scheme. Additionally, we reproduce results from the literature of basis transformations involving the rotation of evanescent operators into the physical basis using our method, without the need to explicitly compute one-loop matrix elements of evanescent operators.
}

\begin{document}
\maketitle

\clearpage

\section{Introduction}
Computations in Effective Field Theories often involve the use of four-dimensional identities in order to reduce Dirac structures. In combination with dimensional regularization such relations have to be generalized to arbitrary space-time dimensions, which is traditionally treated using evanescent operators \cite{Buras:1989xd,Buras:2020xsm,Dugan:1990df}. A simpler way of dealing with such complications has been proposed recently in the literature in the form of one-loop Fierz transformations 
\cite{Aebischer:2022aze,Aebischer:2022rxf}. The traditional Fierz identities \cite{Fierz:1939zz}, which relate combinations of gamma matrices and spinors, are manifestly valid in $d=4$ space-time dimensions. When performing basis transformations at the one-loop level, such identities have to be generalized in order to accommodate the contributions from evanescent operators. Such Fierz-evanescent operators, even though vanishing in $d=4$ space-time dimensions, can generate finite contributions when inserted into divergent loop-diagrams. These finite contributions, which fix the scheme of the calculation \cite{Herrlich:1994kh}, can be computed at any given loop-order for a complete set of four-Fermi operators.
The respective contributions from evanescent operators can then be interpreted as one-loop corrections to the (tree-level) Fierz identities: given an operator $\tilde Q$ and its Fierz-conjugated operator $\mathcal{F}Q$, the traditional way of dealing with Fierz identities in loop calculations is to introduce an evanescent operator $E$, defined by the following equation

\begin{equation}\label{eq:EV}
	\tilde Q=\mathcal{F}Q +E\,.
\end{equation}

\noindent
Although the evanescent operator, $E$, is rank-$\epsilon$ and vanishes in the limit $d\to 4$, divergences in one-loop matrix elements can result in evanescent operators contributing to physical results. 
Therefore, when relating physical quantities between different operator bases, finite loop effects must be taken into account when relating the bases.
In the framework of one-loop Fierz identities, these physical results can be viewed as a correction to the four-dimensional basis change

\begin{equation}\label{eq:1LF}
	\tilde Q=\mathcal{F}Q +\sum_{b}\sum_i\frac{\alpha_b}{4\pi} a^{(b)}_i Q_i\,,
\end{equation}

\noindent
where the $a^{(b)}_i$ denote the scheme-dependent shifts resulting from $\mathcal{O}(\alpha_b$) corrections to the operator and its Fierz-conjugate and $\alpha_b$ denote all possible couplings in the theory. These one-loop shifts are equivalent to the finite contributions resulting from the evanescent operator $E$ in eq.~\eqref{eq:EV}. Note, that the sum in eq.~\eqref{eq:1LF} only runs over the physical operators, $Q_i$, since no Fierz-evanescent operators are introduced.
This novel way of interpreting evanescent contributions as one-loop corrections to Fierz identities facilitates matching computations as well as one-loop basis changes, since Fierz-evanescent operators can be replaced with the generalized relation Eq.~\eqref{eq:1LF} in such calculations. This makes bookkeeping much easier, since the set of relevant evanescent operators is reduced. Moreover, the one-loop shifts can be computed once and for all in a general basis and then reused in future calculations. Such a general computation has been performed for the first time in \cite{Aebischer:2022aze}, where the one-loop QCD and QED shifts to all possible four-Fermi operators were considered. In \cite{Aebischer:2022rxf} the notion of one-loop Fierz identities has been generalized for the first time to include contributions from dipole operators, which arise from penguin diagrams. 
 The framework of one-loop Fierz identities has been proven useful in several NLO computations, such as basis changes and matching computations \cite{Aebischer:2022tvz,Aebischer:2022anv,Aebischer:2018acj,Aebischer:2020dsw,Aebischer:2021raf,Aebischer:2021hws,Fuentes-Martin:2022vvu}. Another virtue of one-loop Fierz identities, which has not been discussed in the literature so far, is the possibility to change between different renormalization schemes simultaneously with a basis change. The key advantage of this procedure is that one only requires the loop-corrected Fierz relations between \textit{physical} operators in the basis; since the evanescent operators are purely scheme-dependent, they may be chosen independently in the respective bases. Moreover, the computation of the loop-corrected Fierz relations does not involve the explicit computation of one-loop matrix elements of evanescent operators, contrary to the standard method of introducing an additional, finite counterterm in a single basis \cite{Gorbahn:2004my, Gorbahn:2009pp}. In this article we explain how to perform such scheme changes in a general manner, using one-loop Fierz identities. Furthermore we provide a proof of the scheme factorization, which lays the foundation for the used method. We illustrate the change of scheme in an explicit example by computing the two-loop anomalous dimension matrix (ADM) relevant for the electric dipole moment (EDM) of the electron. In this example, the renormalization scheme is changed from the naive dimensional regularization (NDR) scheme to the Larin scheme \cite{Larin:1993tq} alongside a transformation to the Fierz-conjugated basis. Finally, we apply our method to the basis change considered in Ref.~\cite{Gorbahn:2009pp} which includes the rotation of evanescent operators into the physical basis. We fully reproduce the results in Ref.~\cite{Gorbahn:2009pp} without the need to explicitly compute one-loop matrix elements of evanescent operators or relate the evanescent operators of the two bases, contrary to the ``finite counterterm'' method of Ref.~\cite{Gorbahn:2009pp}.

The rest of the article is structured as follows: In Sec.~\ref{sec:procedure} the general procedure on how to obtain one-loop shifts for different renormalization schemes is presented. In Sec.~\ref{sec:proof} we provide a proof of the scheme-factorization in one-loop Fierz identities. In Sec.~\ref{sec:ex} we illustrate the advantageous features of our method in several examples. Finally, we conclude in Sec.~\ref{sec:concl}. Useful details concerning the two-loop computation are collected in the Appendix: The greek identities \cite{Tracas:1982gp,Buras:1989xd} used throughout the calculation are collected in App.~\ref{app:greek} and details concerning renormalization are given in App.~\ref{app:ren}.

\section{Procedure}\label{sec:procedure}

In this section we describe how to perform a change of renormalization scheme in the framework of one-loop Fierz identities. We note that the procedure on how to obtain the scheme change between 
two operator bases is very similar to the one discussed in \cite{Aebischer:2022aze}, with the only difference that in this case different schemes are employed. Additionally, we extend the notion
of the one-loop Fierz identities to include basis changes where evanescent and physical operators mix in the basis change; the shifts computed in Ref~\cite{Aebischer:2022aze} must be augmented
by additional finite contributions to account for such cases.

To remedy this, we realize that the shifts computed in \cite{Aebischer:2022aze} arise from relations between \textit{bare} amplitudes computed in dimensional regularization. In this
case, the standard four-dimensional Fierz relations must be extended by $\mathcal{O}(\epsilon)$ contributions in general. Consider an operator basis $\vec{Q}$ and its Fierz-related basis
$\vec{\T{Q}}$ such that in four dimensions, we have
\begin{equation}
	\vec{\T{Q}} \overset{d=4}{=} \mathcal{F}\vec{Q}\,.
\end{equation}
When continuing to $d$-dimensions, this relation no longer necessarily holds, so we must introduce Fierz-evanescent operators into the basis. The form of these operators
will vary depending which basis we are working in, and importantly, the $\mathcal{O}(\epsilon)$ contributions arising in the operator can differ between bases. In general, such
operators can take the form
\begin{equation}
	\vec{E} = K\big(\vec{\T{Q}} - \big(\mathcal{F} + \epsilon \Sigma\big)\vec{Q}\big), \quad \text{or} \quad
	\vec{\T{E}} = \T{K}\big(\vec{Q} - \big(\mathcal{F}^{-1} + \epsilon \T{\Sigma}\big)\vec{\T{Q}}\big)\,,
\end{equation}
where the first operator arises in the $\vec{Q}$ basis and the second in the $\vec{\T{Q}}$ basis. Here, we note that
\begin{equation}
	\T{\Sigma} \neq - \mathcal{F}^{-1}\Sigma\mathcal{F}^{-1}\,,
\end{equation}
in general and the difference amounts to a choice of renormalization scheme. However, this is not the only choice of renormalization scheme available in each basis:
in particular, we also must know how to treat $d$-dimensional Dirac algebra in each basis, e.g. using Naive Dimensional Regularization (NDR), the `t Hooft-Veltmann or the Larin scheme.

As we will prove in the next section, at the NLO level, a general change of basis can be performed with the fully shifted Fierz relation between
four-dimensional physical operators, given by
\begin{equation}
	\vec{\T{Q}}_{\T{\Sigma};\T{S}} = \Big(\mathcal{F} + \Delta\Big)\vec{Q}_{\Sigma;S}\,,
\end{equation}
where the $\T{\Sigma}$ and $\Sigma$ subscripts denote the different evanescent schemes, while $\T{S}$ and $S$ represent different ``Dirac schemes,'' i.e. the scheme chosen
for treating $d$-dimensional Dirac algebra\footnote{Here, the Dirac scheme
also contains the choice of definition of evanescent operators which do not rotate into the physical basis when performing a basis transformation.}. The shift is then computed using
\begin{equation}\label{eq:masterformula}
	\Delta\eval{\vec{Q}}^{(0)} = P_{Q;\Sigma}\Big(\eval{\vec{\T{Q}}}_{\T{\Sigma};\T{S}}^{(1)}\Big)
			- \Big(\mathcal{F} + \epsilon \Sigma\Big)\eval{\vec{Q}}^{(1)}_{\Sigma;S}\,,
\end{equation}
where $\eval{\mathcal{O}}^{(0)}$ is the tree-level matrix element of the operator, $\mathcal{O}$, and $\eval{\mathcal{O}}^{(1)}_{\Sigma;S}$ is the bare one-loop
matrix element of the operator, $\mathcal{O}$, in the evanescent scheme $\Sigma$ and Dirac scheme $S$. The operator $P_{Q;\Sigma}$ denotes the projection of tree-level
matrix elements onto the $\vec{Q}$-basis using the $\Sigma$-scheme, i.e.
\begin{equation}
	P_{Q;\Sigma}\,\mathcal{M}\Big(\eval{\vec{\T{Q}}}^{(0)}\Big) = \mathcal{M}\Big(\big(\mathcal{F} + \epsilon \Sigma\big)\eval{\vec{Q}}^{(0)}\Big)\,.
\end{equation}

\section{Proof of Scheme Factorization}\label{sec:proof}

From Eq.~\eqref{eq:masterformula}, it is clear the $\Delta$ can, in general, depend on \textit{both} renormalization schemes, $\Sigma$ and $\tilde{\Sigma}$ as well as $S$ and $\T{S}$. 
On one hand, we may then worry that, when transforming relevant NLO quantities (e.g. NLO matching conditions for Wilson coefficients or two-loop anomalous dimension matrices)
from the $\vec{Q}$ basis to the $\vec{\tilde{Q}}$ basis, we mix schemes, making it difficult to compare results computed in the 
two separate bases and schemes. On the other hand, one main application of the methodology presented in Section~\ref{sec:procedure} is that it facilitates 
transformations to a new basis where a simpler renormalization scheme can be used. Therefore, for this to be an effective strategy, we must ensure that the double scheme-dependence 
appearing in $\Delta$ factorizes in such a way that one of the two scheme dependencies always cancels in transformations between bases, leaving only the scheme used in the new basis. 
In this section, we prove that the one-loop Fierz identities do indeed satisfy such a factorization of renormalization schemes.

We begin by considering the simpler case where the bases utilize the same renormalization scheme, $(\Sigma;S)$. In this case, we may write the components of Eq.~\eqref{eq:masterformula} as
\begin{equation}
	\begin{split}
		\eval{\vec{\T{Q}}}_{\Sigma;S}^{(1)} &= \Big(r^{(1)}_{\T{Q}\T{Q}} \mathcal{F} + r^{(1)}_{\T{Q}Q} + \epsilon r^{(1)}_{\T{Q}\T{Q}} \Sigma\Big)\eval{\vec{Q}}^{(0)}\,,\\[5pt]
		\eval{\vec{Q}}_{\Sigma;S}^{(1)} &= \Big(r^{(1)}_{QQ} + r^{(1)}_{Q\T{Q}} \mathcal{F} + \epsilon r^{(1)}_{Q\T{Q}} \Sigma\Big)\eval{\vec{Q}}^{(0)}\,.
	\end{split}
\end{equation}
Expanding the one-loop amplitudes by their order in $\epsilon$, i.e.
\begin{equation}
	r_{ij}^{(1)} = \sum_n \frac{1}{\epsilon^n}r^{(1;n)}_{ij}\,,
\end{equation}
this gives the explicit expression for the partial shift
\begin{equation}\label{eq:partialshift}
	\begin{split}
		\hat{\Delta} =& \,r^{(1;0)}_{\T{Q}\T{Q}}\mathcal{F} + r^{(1;0)}_{\T{Q}Q} - \mathcal{F}r^{(1;0)}_{QQ} - \mathcal{F}r^{(1;0)}_{Q\T{Q}}\mathcal{F} \\[5pt]
			& + r^{(1;1)}_{\T{Q}\T{Q}}\Sigma - \Sigma r^{(1;1)}_{QQ} - \Sigma r^{(1;1)}_{Q\T{Q}}\mathcal{F} - \mathcal{F}r^{(1;1)}_{Q \T{Q}}\Sigma\,,
	\end{split}
\end{equation}
where
\begin{equation}
	\hat{\Delta}\eval{\vec{Q}}^{(0)} = \eval{\vec{\T{Q}}}_{\Sigma;S}^{(1)} 
			- \Big(\mathcal{F} + \epsilon \Sigma\Big)\eval{\vec{Q}}^{(1)}_{\Sigma;S}\,.
\end{equation}
Here, we note that the $1/\epsilon$ poles in the amplitudes are scheme-independent, and
therefore must cancel between the two bases when four-dimensional Fierz relations are applied, i.e.
\begin{equation}\label{eq:epidentity}
	r^{(1;1)}_{\T{Q}\T{Q}}\mathcal{F} + r^{(1;1)}_{\T{Q}Q} - \mathcal{F}r^{(1;1)}_{QQ} - \mathcal{F}r^{(1;1)}_{Q\T{Q}}\mathcal{F} = 0\,.
\end{equation}

The result in Eq.~\eqref{eq:partialshift} gives the transformation between operator bases, but does not change schemes. To do this, we must include an additional finite contribution such that
\begin{equation}
	\big(\mathds{1} + \T{\Delta}\mathcal{F}^{-1}\big)\eval{\vec{\T{Q}}}_{\Sigma;S} = \eval{\vec{\T{Q}}}_{\T{\Sigma};\T{S}}\,,
\end{equation}
giving at the NLO level\footnote{No factors of $\T{\Sigma}$ or $\Sigma$ are necessary on the left-hand side of Eq.~\eqref{eq:schemeshift} since the $1/\epsilon$ poles
are scheme-independent. Therefore, the divergences in the one-loop matrix elements in the two schemes are the same and the right-hand side is well-defined in the limit $d\to 4$.}
\begin{equation}\label{eq:schemeshift}
	\T{\Delta}\mathcal{F}^{-1}\eval{\vec{\T{Q}}}^{(0)} = \eval{\vec{\T{Q}}}_{\T{\Sigma};\T{S}}^{(1)} - \eval{\vec{\T{Q}}}_{\Sigma;S}^{(1)}\,.
\end{equation}
To perform this calculation, we first note that in the $\vec{\T{Q}}$ basis, the $\vec{Q}$ operators are given in $d$-dimensions by
\begin{equation}
	\vec{Q}_{\Sigma} = \Big(\mathcal{F}^{-1} - \epsilon \mathcal{F}^{-1}\Sigma\mathcal{F}^{-1}\Big)\vec{\T{Q}}\,, \quad
	\vec{Q}_{\T{\Sigma}} = \Big(\mathcal{F}^{-1} + \epsilon \T{\Sigma}\Big)\vec{\T{Q}}\,,
\end{equation}
in the $\Sigma$ and $\T{\Sigma}$ evanescent schemes, respectively. Then, the one-loop matrix elements are given by
\begin{equation}\label{eq:tildbasis}
	\begin{split}
		\eval{\vec{\T{Q}}}^{(1)}_{\Sigma;S} =& \Big(r_{\T{Q}\T{Q}}^{(1)} + r_{\T{Q}Q}^{(1)}\mathcal{F}^{-1}
				- \epsilon r_{\T{Q}Q}^{(1)}\mathcal{F}^{-1}\Sigma\mathcal{F}^{-1}\Big)\eval{\vec{\T{Q}}}^{(0)}\,, \\[5pt]
		\eval{\vec{\T{Q}}}^{(1)}_{\T{\Sigma};\T{S}} =& \Big(\T{r}_{\T{Q}\T{Q}}^{(1)} + \T{r}_{\T{Q}Q}^{(1)}\mathcal{F}^{-1}
				+ \epsilon \T{r}_{\T{Q}Q}^{(1)}\T{\Sigma}\Big)\eval{\vec{\T{Q}}}^{(0)}\,,
	\end{split}
\end{equation}
which gives
\begin{equation}\label{eq:deltild}
	\T{\Delta}\mathcal{F}^{-1} = \T{r}_{\T{Q}\T{Q}}^{(1;0)} + \T{r}_{\T{Q}Q}^{(1;0)}\mathcal{F}^{-1} - r_{\T{Q}\T{Q}}^{(1;0)} - r_{\T{Q}Q}^{(1;0)}\mathcal{F}^{-1}
		+ r_{\T{Q}Q}^{(1;1)}\T{\Sigma} + r_{\T{Q}Q}^{(1;1)}\mathcal{F}^{-1}\Sigma\mathcal{F}^{-1}\,,
\end{equation}
where the first four terms change from Dirac scheme $S$ to $\T{S}$ and the last two terms change from evanescent scheme $\Sigma$ to $\T{\Sigma}$.
We have not differentiated between the two schemes for the $1/\epsilon$ pieces of the amplitudes as these are scheme-independent (i.e. $\T{r}^{(1;1)}=r^{(1;1)})$.
Combining these two contributions to the transformation, $\Delta = \hat{\Delta} + \T{\Delta}$, we arrive exactly at Eq.~\eqref{eq:masterformula}.

Note, we can arrive directly to the same definition of $\Delta$ using Eq.~\eqref{eq:tildbasis} along with the projector, $P_{Q;\Sigma}$ to find
\begin{equation}
	P_{Q;\Sigma}\Big(\eval{\vec{\T{Q}}}^{(1)}_{\T{\Sigma};\T{S}}\Big) = \Big(\T{r}_{\T{Q}\T{Q}}^{(1)}\mathcal{F}
	+ \T{r}_{\T{Q}Q}^{(1)} + \epsilon\T{r}^{(1)}_{\T{Q}\T{Q}}\Sigma + \epsilon\T{r}_{\T{Q}Q}^{(1)}\mathcal{F}^{-1}\Sigma
	+ \epsilon \T{r}^{(1)}_{\T{Q}Q}\T{\Sigma}\mathcal{F}\Big)\eval{\vec{Q}}^{(0)}\,.
\end{equation}
Using this in Eq.~\eqref{eq:masterformula} gives exactly $\Delta = \hat{\Delta} + \T{\Delta}$.

Next, we will prove that this is equivalent to the standard methods used for higher-order basis transformations (see e.g. Ref~\cite{Gorbahn:2004my}). The tree-level basis transformation using this technique is
given by
\begin{equation}\label{eq:basischange}
	\vec{\T{Q}} = R\big(\vec{Q} + W \vec{E}\big), \quad \vec{\T{E}} = M\big(\epsilon U\vec{Q} + \big(\mathds{1} + \epsilon U W\big)\vec{E}\big)\,.
\end{equation}
By inspection, the matrices used in Eq.~\eqref{eq:basischange} are given in terms of matrices already introduced as
\begin{equation}\label{eq:matrixrels}
	\begin{split}
		R &= \mathcal{F}+\epsilon \Sigma\,, \quad W = \mathcal{F}^{-1} K^{-1}-\epsilon\mathcal{F}^{-1} \Sigma\mathcal{F}^{-1}K^{-1}\,, \\ 
		M &= - \T{K}\mathcal{F}^{-1}K^{-1}+\epsilon\T{K}\mathcal{F}^{-1} \Sigma\mathcal{F}^{-1}K^{-1}\,, \quad U = K\Sigma + K \mathcal{F} \T{\Sigma}\mathcal{F}\,.
	\end{split}
\end{equation}
The basis change is carried out by adding an additional finite counterterm $Z_{\T{Q}\T{Q}}^{(1;0)}$ to the physical operators in the new basis which can be computed
using\footnote{Since the counterterm in Eq.~\eqref{eq:finitecounterterm} is finite, the $O(\epsilon)$ pieces of the matrices in Eq.~\eqref{eq:matrixrels} are irrelevant.}
\begin{equation}\label{eq:finitecounterterm}
	Z_{\T{Q}\T{Q}}^{(1;0)} = R\Big[W Z_{EQ}^{(1;0)} - \Big(Z_{QE}^{(1;1)} + W Z_{EE}^{(1;1)} - Z_{QQ}^{(1;1)}W\Big)U\Big]R^{-1}\,.
\end{equation}
The counterterms on the right-hand side of Eq.~\eqref{eq:finitecounterterm} can be found by inserting physical and evanescent operators into
one-loop diagrams using the $(\Sigma;S)$ scheme. They are given by
\begin{equation}
	\begin{split}
		&Z_{QQ}^{(1;1)} = - r_{QQ}^{(1;1)} - r_{Q\T{Q}}^{(1;1)}\mathcal{F}, \quad Z_{QE}^{(1;1)} = - r_{Q\T{Q}}^{(1;1)}K^{-1}, \\[5pt]
		&Z_{EE}^{(1;1)} = - K r_{\T{Q}\T{Q}}^{(1;1)}K^{-1} + K \mathcal{F} r_{Q\T{Q}}^{(1;1)}K^{-1}\,, \\[5pt]
		&Z_{EQ}^{(1;0)} = - K r_{\T{Q} Q}^{(1;0)} - K r_{\T{Q}\T{Q}}^{(1;0)} \mathcal{F} 
			+ K \mathcal{F} r_{QQ}^{(1;0)} + K \mathcal{F}r_{Q\T{Q}}^{(1;0)}\mathcal{F} \\[5pt]
			&\hspace{2cm}- K r_{\T{Q}\T{Q}}^{(1;1)}\Sigma + K \mathcal{F}r_{Q\T{Q}}^{(1;1)}\Sigma
			+ K\Sigma r_{Q\T{Q}}^{(1;1)}\mathcal{F} + K \Sigma r_{QQ}^{(1;1)}\,.
	\end{split}
\end{equation}
Using these in Eq.~\eqref{eq:finitecounterterm} and using Eq.~\eqref{eq:epidentity}, we find
\begin{equation}
	\begin{split}
		Z_{\T{Q}\T{Q}}^{(1;0)} = - \hat{\Delta}\mathcal{F}^{-1} - r_{\T{Q}Q}^{(1;1)}\T{\Sigma}
		- r_{\T{Q}Q}^{(1;1)}\mathcal{F}^{-1}\Sigma\mathcal{F}^{-1}\,.
	\end{split}
\end{equation}
This transformation changes the basis $\vec{Q}\to\vec{\T{Q}}$ as well as the evanescent scheme, $\Sigma\to\T{\Sigma}$, but does not change the Dirac scheme. This is
accomplished by another finite counterterm
\begin{equation}
	\T{Z}_{QQ}^{(1;0)} = - \T{r}_{\T{Q}\T{Q}}^{(1;0)} - \T{r}_{\T{Q}Q}^{(1;0)}\mathcal{F}^{-1} + r_{\T{Q}\T{Q}}^{(1;0)} + r_{\T{Q}Q}^{(1;0)}\mathcal{F}^{-1}\,,
\end{equation}
from which it is clear that\footnote{The relative sign between $Z_{\T{Q}\T{Q}}$ and
$\Delta$ comes from the fact that $Z_{\T{Q}\T{Q}}$ is a counterterm associated with the Wilson coefficient, not the operator.} 
\begin{equation}
	Z_{QQ}^{(1;0)} + \T{Z}_{QQ}^{(1;0)} = - \Delta\mathcal{F}^{-1}\,,
\end{equation}
and the two methods are equivalent. Therefore, the ``shift method'' properly treats both the basis transformation \textit{and} the scheme transformation simultaneously.

We reiterate that Eq.~\eqref{eq:masterformula} relates only one-loop matrix elements of physical operators in the two bases; no one-loop matrix elements of evanescent operators need be computed. 
Additionally, each operator loop-insertion is computed in its respective scheme with no further scheme transformation necessary. It therefore becomes unnecessary to relate the evanescent operators 
between the two bases/schemes as must be done in general in the traditional method, i.e. Eq.~\eqref{eq:basischange}.

\section{Examples}\label{sec:ex}

\subsection{Example One: BSM Electron EDM}

As an example of utilizing the scheme factorization of the Fierz procedure, we examine the calculation performed in Ref.~\cite{Brod:2023wsh}. The authors of~\cite{Brod:2023wsh} calculate the contributions to the electron EDM arising from NLO QCD and QED
running below the electroweak scale, assuming Beyond the Standard Model (BSM) pseudoscalar couplings between the heavy bottom and charm quarks and the Higgs. For the sake of simplicity, we will focus only on the bottom quark case in this example. The calculation is performed in the basis well-suited for the matching calculation
\begin{equation}
	\mathcal{L}_{\text{Lar}} = -\sqrt{2}G_F\Big(C_1^{eb}\mathcal{O}_1^{eb} + C_1^{be}\mathcal{O}_1^{be} 
		+ C_2\mathcal{O}_2 + C_3\mathcal{O}_3\Big)\,,
\end{equation}
with the Fermi constant $G_F$, and where the fields and operators are defined by $\psi_{i,j} = e$ or $b$,
\begin{equation}\label{eq:lar_defs}
	\mathcal{O}_1^{ij} = \big(\bar{\psi}_i\psi_i\big)\big(\bar{\psi}_j\,i \gamma_5\,\psi_j\big), \quad
	\mathcal{O}_2 = \frac{1}{2}\epsilon^{\mu \nu \rho \sigma}\big(\bar{e}\sigma_{\mu \nu}e\big)\big(\bar{b}\sigma_{\rho\sigma}b\big), \quad
	\mathcal{O}_3 = \frac{Q_e m_b}{2 e}\big(\bar{e}\sigma^{\mu \nu}e\big)\tilde{F}_{\mu \nu}\,.
\end{equation}
In Eq.~\eqref{eq:lar_defs}, the antisymmetric tensors are defined by $\sigma^{\mu \nu} = \frac{i}{2}[\gamma^\mu,\gamma^\nu]$ and
$\tilde{F}_{\mu\nu} = \frac{1}{2}\,\epsilon_{\mu \nu \rho \sigma} F^{\rho \sigma}$. While this basis is useful for simplifying the matching
calculation, it adds an additional difficulty when computing the anomalous dimension matrix due to the fact that it introduces traces over
spin indices which include factors of $\gamma_5$. Such traces require the use of a non-trivial renormalization scheme when calculating the mixing into
the dipole operator. In Ref.~\cite{Brod:2023wsh}, the Larin scheme is used, where $\gamma_5$ is replaced in favor of its definition in terms of the well-defined
$d$-dimensional gamma-matrices\footnote{For an extensive introduction to Renormalization schemes with non-anticommuting $\gamma_5$ see \cite{Belusca-Maito:2023wah}.}
\begin{equation}
	\gamma_5 = \frac{i}{4!}\epsilon_{\mu \nu \rho \sigma}\gamma^\mu \gamma^\nu \gamma^\rho \gamma^\sigma\,.
\end{equation}

We can, however, choose a different basis which makes the calculation of the ADM far simpler. For this, we choose
\begin{equation}
	\mathcal{L}_{\text{NDR}} = -\sqrt{2}G_F\Big(\T{C}_s \T{\mathcal{O}}_s + \T{C}_v \T{\mathcal{O}}_v
		+ \T{C}_t \T{\mathcal{O}}_t + \T{C}_3 \T{\mathcal{O}}_3\Big)\,,
\end{equation}
where, suppressing color indices,
\begin{equation}\label{eq:PHYSbasis}
	\begin{split}
		\T{\mathcal{O}}_s =& \frac{1}{2}\Big[\big(\bar{b}i\gamma_5 e\big)\big(\bar{e} b\big) 
			+ \big(\bar{b} e\big)\big(\bar{e}i\gamma_5 b\big)\Big]\,, \quad
		\T{\mathcal{O}}_v = \frac{1}{2}\Big[\big(\bar{b}i\gamma^\mu \gamma_5 e\big)\big(\bar{e}\gamma_\mu b\big) 
			- \big(\bar{b}\gamma^\mu e\big)\big(\bar{e}i\gamma_\mu \gamma_5 b\big)\Big]\,, \\[0.5em]
		\T{\mathcal{O}}_t =& \frac{1}{2}\Big[\big(\bar{b}i\sigma_{\mu \nu}\gamma_5 e\big)\big(\bar{e}\sigma^{\mu \nu} b\big) 
			+ \big(\bar{b}\sigma_{\mu \nu} e\big)\big(\bar{e}i\sigma^{\mu \nu}\gamma_5 b\big)\Big]\,, \quad
		\T{\mathcal{O}}_3 = \frac{Q_e m_b}{2e}\big(\bar{e}i\sigma_{\mu \nu}\gamma_5 e\big) F^{\mu \nu}\,.
	\end{split}
\end{equation}
This basis is particularly convenient due to the fact that it allows for the mixing of the four-quark operators into the dipole operators to
be computed using NDR, both simplifying the evaluation of individual diagrams and reducing the number
of necessary unphysical operators. Denoting the vectors of operators in the original and new basis as 
$\mathcal{O}_i = (\mathcal{O}^{eb}_1,\mathcal{O}^{be}_1,\mathcal{O}_2,\mathcal{O}_3)$ and
$\T{\mathcal{O}}_i = (\T{\mathcal{O}}_s, \T{\mathcal{O}}_v, \T{\mathcal{O}}_t, \T{\mathcal{O}}_3)$, respectively, the tree-level basis
transformation\footnote{We have used the relation $\sigma^{\mu\nu}\gamma_5=-\frac{i}{2}\epsilon^{\mu\nu\rho\sigma}\sigma_{\rho\sigma}$.}, $\T{\mathcal{O}}_i = \mathcal{F}_{ij}\mathcal{O}_j$ is given by
\begin{equation}
	\mathcal{F}_{ij} = \begin{pmatrix}
		-\frac{1}{4} & -\frac{1}{4} & -\frac{1}{8} & 0 \\[0.5em]
		1 & -1 & 0 & 0 \\[0.5em]
		-3 & -3 & \frac{1}{2} & 0 \\[0.5em]
		0 & 0 & 0 & 1
	\end{pmatrix}\,.
\end{equation}

\begin{figure}[tb]
	\centering
	\begin{tikzpicture}
	\begin{feynman}
		\vertex (i4) {\(\bar{b}\)};
		\vertex[below = 1.5cm of i4] (fake1);
		\node[left = 1.5cm of fake1, crossed dot] (j2);
		\vertex[below = 3cm of i4] (i3) {\(e\)};
		\vertex[left = 3cm of i4] (i1) {\(b\)};
		\vertex[below = 3cm of i1] (i2) {\(\bar{e}\)};
		\vertex[below = 0.5cm of i1] (fake3);
		\vertex[right = 0.45cm of fake3] (a);
		\vertex[below = 0.5cm of i4] (fake4);
		\vertex[left = 0.55cm of fake4] (b);
		\vertex[below = 0.75cm of j2] (label) {\hspace{0.2cm}\(\mathcal{O}_i\)};
		\diagram*{
			(i1) --[fermion] (j2) --[fermion] (i2),
			(i3) --[fermion] (j2) --[fermion] (i4),
			(a) --[gluon, in = 160, out = 25] (b)
		};
	\end{feynman}
	\end{tikzpicture}
	\caption{Feynman diagram contributing to the one-loop QCD ADM with single
	operator insertion, $\mathcal{O}_i$, where $i = s, v, t$.}
	\label{fig:oneloop}
\end{figure}
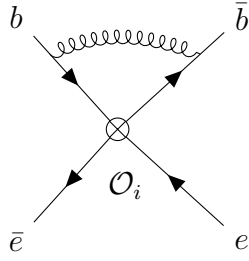

At the one-loop level, there are only two relevant evanescent operators into which the physical operators mix:
\begin{equation}\label{eq:EVsNDR}
	\begin{split}
		\T{E}_v =& \frac{1}{2}\Big[\big(\bar{b}\gamma^\mu \gamma^\nu \gamma^\rho i\gamma_5\,e\big)
			\big(\bar{e}\gamma_\mu \gamma_\nu \gamma_\rho\,b\big)
			- \big(\bar{b}\gamma^\mu \gamma^\nu \gamma^\rho\,e\big)
			\big(\bar{e}\gamma_\mu \gamma_\nu \gamma_\rho i\gamma_5\,b\big)\Big]
			- (4 + a_v \epsilon)\T{\mathcal{O}}_v\,, \\[0.5em]
		\T{E}_t =& \frac{1}{2}\Big[\big(\bar{b}\gamma^\mu \gamma^\nu \gamma^\rho \gamma^\sigma i\gamma_5\,e\big)
			\big(\bar{e}\gamma_\mu \gamma_\nu \gamma_\rho \gamma_\sigma\,b\big)
			+ \big(\bar{b}\gamma^\mu \gamma^\nu \gamma^\rho \gamma^\sigma\,e\big)
			\big(\bar{e}\gamma_\mu \gamma_\nu \gamma_\rho \gamma_\sigma i\gamma_5\,b\big)\Big]\\[0.5em]
			& \quad - (64 + a_s \epsilon)\T{\mathcal{O}}_s + (16 + a_t\epsilon)\T{\mathcal{O}}_t\,,
	\end{split}
\end{equation}
where $a_s$, $a_v$, and $a_t$ parameterize the renormalization scheme dependence in the tilded basis. 
Having the complete NDR operator basis at hand, one can perform the full one- and two-loop computation of the ADMs in this basis. For this purpose we have used the methods outlined in \cite{Chetyrkin:1997fm} in self-written \texttt{Mathematica} code, cross-referenced with the \texttt{MaRTIn} \texttt{FORM} routines~\cite{Brod:2024zaz}. At one-loop QCD, there is only one Feynman diagram with a single operator insertion which contributes to the ADM, shown in Figure~\ref{fig:oneloop}. Examples of two-loop diagrams with single operator insertions are shown in Figure~\ref{fig:twoloop}.

At one-loop order in QCD one finds the following (scheme-independent) ADM\footnote{In the following we have set $n_f=5$ and $N_c = C_A = 3$.}

\begin{equation}
	\T{\gamma}^{(0),\text{SI}} = \begin{pmatrix}
		0 & 0 & -\frac{2}{3} & 0 \\[0.5em]
		0 & -8 & 0 & 0 \\[0.5em]
		-32 & 0 & -\frac{16}{3} & 0 \\[0.5em]
		0 & 0 & 0 & 8
	\end{pmatrix}\,.
\end{equation}

The two-loop ADM is scheme-dependent, due to the choice of evanescent operators in Eq.~\eqref{eq:EVsNDR}. One can therefore split the two-loop QCD anomalous dimension matrix into a scheme-independent and three scheme-dependent parts in the following way:

\begin{equation}
	\tilde\gamma^{(1)} = \tilde\gamma^{(1),\text{SI}} + a_s \tilde\gamma^{(1),s} + a_v \tilde\gamma^{(1),v} + a_t \tilde\gamma^{(1),t}\,,
\end{equation}

\noindent where the individual parts are given by 

\begin{equation}
	\begin{split}
		\T{\gamma}^{(1),\text{SI}} =& \begin{pmatrix}
			\frac{284}{9} & 0 & -\frac{275}{27} & 0 \\[0.5em]
			0 & -92 & 0 & 0 \\[0.5em]
			48 & 0 & -\frac{3340}{27} & 0 \\[0.5em]
			0 & 0 & 0 & \frac{1012}{9}
		\end{pmatrix}\,, \qquad
		\T{\gamma}^{(1),s} = \begin{pmatrix}
			\frac{2}{9} & 0 & 0 & 0 \\[0.5em]
			0 & 0 & 0 & 0 \\[0.5em]
			\frac{62}{9} & 0 & -\frac{2}{9} & 0 \\[0.5em]
			0 & 0 & 0 & 0
		\end{pmatrix}\,,\\[1.0em]
		\T{\gamma}^{(1),v} =& \begin{pmatrix}
			0 & 0 & 0 & 0 \\[0.5em]
			0 & -\frac{46}{9} & 0 & 0 \\[0.5em]
			0 & 0 & 0 & 0 \\[0.5em]
			0 & 0 & 0 & 0
		\end{pmatrix}\,, \qquad
		\T{\gamma}^{(1),t} = \begin{pmatrix}
			0 & 0 & -\frac{2}{9} & 0 \\[0.5em]
			0 & 0 & 0 & 0 \\[0.5em]
			\frac{32}{3} & 0 & -\frac{46}{9} & 0 \\[0.5em]
			0 & 0 & 0 & 0
		\end{pmatrix}\,.
	\end{split}
\end{equation}

 We reiterate that the computation in NDR is much simpler compared to the Larin scheme, partially exemplified by the reduction of the number of evanescent operators from nine in Ref.~\cite{Brod:2023wsh} to two. Additionally, Dirac algebra manipulations are far simpler in NDR, as is the projection of matrix elements onto tree-level operators.

 \begin{figure}[t]
	\centering
	\begin{tikzpicture}
	\begin{feynman}
		\vertex (i4) {\(\bar{b}\)};
		\vertex[below = 1.5cm of i4] (fake1);
		\node[left = 1.5cm of fake1, crossed dot] (j2);
		\vertex[below = 3cm of i4] (i3) {\(e\)};
		\vertex[left = 3cm of i4] (i1) {\(b\)};
		\vertex[below = 3cm of i1] (i2) {\(\bar{e}\)};
		\vertex[below = 0.5cm of i1] (fake3);
		\vertex[right = 0.45cm of fake3] (a);
		\vertex[below = 0.5cm of i4] (fake4);
		\vertex[left = 0.55cm of fake4] (b);
		\vertex[below = 0.75cm of j2] (label) {\hspace{0.2cm}\(\mathcal{O}_i\)};
		\vertex[above = 1.15cm of j2] (fake2);
		\vertex[left = 0.25cm of fake2] (c);
		\vertex[right = 0.35cm of fake2] (d);
		\diagram*{
			(i1) --[fermion] (j2) --[fermion] (i2),
			(i3) --[fermion] (j2) --[fermion] (i4),
			(a) --[gluon] (c) --[fermion, half left] (d)--[gluon] (b),
			(d) --[fermion, half left] (c)
		};
	\end{feynman}
	\end{tikzpicture}
	\hspace{0.5cm}
	\begin{tikzpicture}
	\begin{feynman}
		\vertex (i4) {\(\bar{b}\)};
		\vertex[below = 1.5cm of i4] (fake1);
		\node[left = 1.5cm of fake1, crossed dot] (j2);
		\vertex[below = 3cm of i4] (i3) {\(e\)};
		\vertex[left = 3cm of i4] (i1) {\(b\)};
		\vertex[below = 3cm of i1] (i2) {\(\bar{e}\)};
		\vertex[below = 0.5cm of i1] (fake3);
		\vertex[right = 0.45cm of fake3] (a);
		\vertex[below = 0.5cm of i4] (fake4);
		\vertex[left = 0.55cm of fake4] (b);
		\vertex[below = 0.75cm of j2] (label) {\hspace{0.2cm}\(\mathcal{O}_i\)};
		\vertex[above = 1.3cm of j2] (c);
		\vertex[above = 0.35cm of j2] (fake2);
		\vertex[right = 0.4cm of fake2] (d);
		\diagram*{
			(i1) --[fermion] (j2) --[fermion] (i2),
			(i3) --[fermion] (j2) --[fermion] (i4),
			(a) --[gluon, in = 180] (c) --[gluon, out = 0] (b),
			(c) --[gluon, out = 270] (d)
		};
	\end{feynman}
	\end{tikzpicture}
	\hspace{0.5cm}
	\begin{tikzpicture}
	\begin{feynman}
		\vertex (i4) {\(\bar{b}\)};
		\vertex[left = 4cm of i4] (i3) {\(b\)};
		\vertex[left = 1.3333cm of i4] (j1);
		\vertex[left = 1.3333cm of j1] (j2);
		\vertex[below = 1.0cm of j1] (j3);
		\vertex[below = 1.0cm of j2] (j4);
		\vertex[below = 2cm of i4] (i2) {\(\bar{e}\)};
		\vertex[below = 2cm of i3] (i1) {\(e\)};
		\node[left = 2cm of i2, crossed dot] (j5);
		\vertex[below = 0.4cm of j5] (label) {\hspace{0.2cm}\(\mathcal{O}_i\)};
		\vertex[above = 0.6cm of j5] (label2) {\(b\)};
		\diagram*{
			(i3) --[fermion] (j2) --[fermion] (j1) --[fermion] (i4),
			(j2) --[gluon] (j4),
			(j1) --[gluon] (j3),
			(i1) --[fermion] (j5) --[fermion] (j4) --[fermion] (j3) --[fermion] (j5) --[fermion] (i2)
		};
	\end{feynman}
	\end{tikzpicture}
	\caption{Representative Feynman diagrams contributing to the two-loop QCD ADM with single
	operator insertion, $\mathcal{O}_i$, where $i = s, v, t$.}
	\label{fig:twoloop}
\end{figure}

Changing now the obtained ADMs back to the matching basis defined in Eq.~\eqref{eq:lar_defs} and the Larin scheme, the one-loop QCD shifts for the involved operators have to be computed. We write the general transformation between the untilded to the tilded basis in the following form:

\begin{equation}
	\T{\mathcal{O}}_i = \Bigg[\mathcal{F}_{ij} 
		+ \sum_{n}\Big(\frac{\alpha_s}{4\pi}\Big)^n\Delta^{(n)}_{ij}\Bigg]\mathcal{O}_j\,.
\end{equation}

At one loop, the shifts, $\Delta$, can be split up again according to their scheme-dependence

\begin{equation}
	\Delta^{(1)} = \Delta^{(1),\text{SI}} + a_s\Delta^{(1),s} + a_v \Delta^{(1),v} + a_t \Delta^{(1),t}\,.
\end{equation}

Following the procedure laid out in Sec.~\ref{sec:procedure} we find the following one-loop QCD shifts, which transform from the basis in Eq.~\eqref{eq:lar_defs} to the basis in Eq.~\eqref{eq:PHYSbasis} and simultaneously from the NDR to the Larin scheme (using the definition of evanescent operators given in \cite{Brod:2023wsh}):

\begin{equation}
	\begin{split}
		\Delta^{(1),\text{SI}} =& \begin{pmatrix}
			\frac{3}{2} & -\frac{7}{6} & \frac{1}{12} & 0 \\[0.5em]
			-\frac{28}{3} & -\frac{4}{3} & 0 & 0 \\[0.5em]
			\frac{110}{3} & \frac{14}{3} & -\frac{13}{3} & 0 \\[0.5em]
			0 & 0 & 0 & 0
		\end{pmatrix}\,, \qquad
		\Delta^{(1),s} = \begin{pmatrix}
			0 & 0 & 0 & 0 \\[0.5em]
			0 & 0 & 0 & 0 \\[0.5em]
			-\frac{1}{12} & -\frac{1}{12} & -\frac{1}{24} & 0 \\[0.5em]
			0 & 0 & 0 & 0
		\end{pmatrix}\,,\\[1.0em]
		\Delta^{(1),v} =& \begin{pmatrix}
			0 & 0 & 0 & 0 \\[0.5em]
			-\frac{1}{3} & \frac{1}{3} & 0 & 0 \\[0.5em]
			0 & 0 & 0 & 0 \\[0.5em]
			0 & 0 & 0 & 0
		\end{pmatrix}\,, \qquad
		\Delta^{(1),t} = \begin{pmatrix}
			0 & 0 & 0 & 0 \\[0.5em]
			0 & 0 & 0 & 0 \\[0.5em]
			1 & 1 & -\frac{1}{6} & 0 \\[0.5em]
			0 & 0 & 0 & 0
		\end{pmatrix}\,.
	\end{split}
\end{equation}

Combining these results together with the ADMs, using the transformation
\begin{equation}
        \tilde{\gamma}^{(1)}_{ij} = \mathcal{F}_{ik}\Big[\gamma_{k\ell}^{(1)} + \mathcal{F}^{-1}_{km}\Delta_{mn}\gamma^{(0)}_{n\ell}
                - \gamma^{(0)}_{km}\mathcal{F}^{-1}_{mn}\Delta_{n\ell} 
		- 2\beta^{(0)}\mathcal{F}^{-1}_{km}\Delta_{m\ell}\Big]\mathcal{F}_{\ell j}^{-1}\,,
\end{equation}
with $\beta^{(0)} = \frac{11}{3}N_c - \frac{2}{3}n_f$ the one-loop QCD beta function, we find the same
ADM as presented in Ref.~\cite{Brod:2023wsh}.\footnote{One of the authors also computed the two-loop mixed QCD-QED, as well as the pure two-loop QED ADMs, together with the corresponding shifts, and again found full agreement with the results in \cite{Brod:2023wsh}, as well as explicit cancellation of the NDR scheme-dependence.}  The resulting ADM in the Larin scheme does not depend on the scheme-dependent coefficients $a_s\,,a_v\,,a_t$ as expected.

\subsection{Example Two: Basis Change in a General Scheme}

Here, we consider a generalization of the change of basis presented in Ref.~\cite{Gorbahn:2009pp}. In Ref.~\cite{Aebischer:2022aze} it was shown that the shift procedure used in the 
reference can exactly reproduce such a change of basis given a modification by ``$U$-terms'' in Eq.~\eqref{eq:finitecounterterm}. We will show that the generalized procedure, Eq.~\eqref{eq:masterformula}, 
needs no further modification and immediately reproduces the proper result.

We begin with the basis defined by
\begin{equation}
	\begin{split}
		&Q_1 = \big(\bar{b}_R\,q_L\big)\big(\bar{b}_R\,q_L\big)\,, \\[5pt]
		&Q_2 = - \big(\bar{b}_R\,\sigma^{\mu\nu}q_L\big)\big(\bar{b}_R\,\sigma_{\mu\nu}q_L\big)\,, \\[5pt]
		&E_1 = \big(\bar{b}_R^i \,q_L^j\big)\big(\bar{b}_R^j\,q_L^i\big) + \Big(\frac{1}{2} - \epsilon \sigma_1\Big)Q_1
			- \Big(\frac{1}{8} - \epsilon \sigma_2\Big)Q_2\,, \\[5pt]
		&E_2 = - \big(\bar{b}_R^i\,\sigma^{\mu\nu}q_L^j\big)\big(\bar{b}_R^j\,\sigma_{\mu\nu} q_L^i\big)
			- (6 - \epsilon \sigma_3)Q_1 - \Big(\frac{1}{2} - \epsilon \sigma_4\Big)Q_2\,,
	\end{split}
\end{equation}
where $\sigma_i$ are unfixed scheme constants and non-trivial color contractions are shown explicitly. We now wish to compute the NLO transformation to the new basis
\begin{equation}
	\begin{split}
		&\T{Q}_1 = \big(\bar{b}_R\,q_L\big)\big(\bar{b}_R\,q_L\big)\,, \\[5pt]
		&\T{Q}_2 = \big(\bar{b}_R^i \,q_L^j\big)\big(\bar{b}_R^j\,q_L^i\big)\,, \\[5pt]
		&\T{E}_1 = \big(\bar{b}_R\,\sigma^{\mu\nu}q_L\big)\big(\bar{b}_R\,\sigma_{\mu\nu}q_L\big)
			+ (4 - \epsilon\T{\sigma}_1)\T{Q}_1 + (8 - \epsilon \T{\sigma}_2)\T{Q}_2\,, \\[5pt]
		&\T{E}_2 = \big(\bar{b}_R^i\,\sigma^{\mu\nu}q_L^j\big)\big(\bar{b}_R^j\,\sigma_{\mu\nu} q_L^i\big)
			+ (4 -\epsilon\T{\sigma}_3)\T{Q}_2 + (8 - \epsilon \T{\sigma}_4)\T{Q}_1\,.
	\end{split}
\end{equation}
We begin by using the ``traditional method'' to compute the finite counterterm $Z_{\T{Q}\T{Q}}^{(1;0)}$ used in the basis transformation. The $O(1/\epsilon)$ counterterms along with the matrices 
$R = \mathcal{F}$, $W$, and $M$ are scheme-independent and are given in \cite{Gorbahn:2009pp}. For the finite evanescent-to-physical counterterm in the original basis, we find
\begin{equation}
	Z_{EQ}^{(1;0)} = \begin{pmatrix}
		\frac{37}{12} + 7\sigma_1 - 20\sigma_2 - \frac{11}{6}\sigma_3 &&& \frac{29}{48} - \frac{1}{12}\sigma_1 + \frac{11}{3}\sigma_2 - \frac{11}{6}\sigma_4 \\[10pt]
		-\frac{73}{3} - \frac{16}{3}\sigma_1 + \frac{29}{3}\sigma_3 - 20\sigma_4 &&& - \frac{1}{12} + \frac{16}{3}\sigma_2 + \frac{1}{12}\sigma_3 + \frac{61}{3}\sigma_4
	\end{pmatrix}\,,
\end{equation}
and the $U$ matrix is given by
\begin{equation}
	U = \begin{pmatrix}
		\sigma_1 - \frac{1}{8}\T{\sigma}_1 + \frac{1}{16}\T{\sigma}_2 &&& - \sigma_2 - \frac{1}{64}\T{\sigma}_2 \\[10pt]
		- \sigma_3 - \frac{1}{2}\T{\sigma}_1 + \frac{1}{4}\T{\sigma}_2 - \frac{1}{2}\T{\sigma}_3 + \T{\sigma}_4 &&&
		- \sigma_4 - \frac{1}{16}\T{\sigma}_2 + \frac{1}{8}\T{\sigma}_3
	\end{pmatrix}\,.
\end{equation}
Using Eq.~\eqref{eq:finitecounterterm}, these results give the finite counterterm
\begin{equation}\label{eq:NiersteCT}
	\begin{split}
		&Z_{\T{Q}_1\T{Q}_1}^{(1;1)} = \frac{2}{3}\sigma_1 - \frac{8}{3}\sigma_2 + \frac{1}{2}\sigma_3 + 2\sigma_4 + \frac{1}{6}\T{\sigma}_1 - \frac{1}{2}\T{\sigma}_4\,,\\[5pt]
		&Z_{\T{Q}_1\T{Q}_2}^{(1;1)} = -\frac{16}{3}\sigma_2 + 4\sigma_4 + \frac{1}{6}\T{\sigma}_2 - \frac{1}{2}\T{\sigma}_3\,,\\[5pt]
		&Z_{\T{Q}_2\T{Q}_1}^{(1;1)} = \frac{11}{2} + 11\sigma_1 - \frac{68}{3}\sigma_2 - \frac{5}{4}\sigma_3 - 5\sigma_4 - \frac{1}{4}\T{\sigma}_1 - \frac{7}{12}\T{\sigma}_4\,,\\[5pt]
		&Z_{\T{Q}_2\T{Q}_2}^{(1;1)} = \frac{29}{6} - \frac{2}{3}\sigma_1 - \frac{16}{3}\sigma_2 - 10\sigma_4 - \frac{1}{4}\T{\sigma}_2 - \frac{7}{12}\T{\sigma}_3\,,
	\end{split}
\end{equation}
which reproduces the results in \cite{Gorbahn:2009pp} for $\sigma_i = 0$, $\T{\sigma}_1 = \T{\sigma}_3 = 2$, and $\T{\sigma}_2 = \T{\sigma}_4 = 8$.

Using the shift method, we begin by inserting physical operators into one-loop QCD diagrams in their respective schemes using the evanescent operators defined above. The four operator insertions are
given by
\begin{equation}
	\begin{split}	
		&\frac{4\pi}{\alpha_s}\eval{Q_1}^{(1)}_{\Sigma;S} = \Big(\frac{1}{\epsilon}\Big\{5 + \frac{8}{3}\xi\Big\} - \frac{76}{9} - \frac{20}{9}\xi + \frac{1}{2}\sigma_3\Big)\eval{Q_1}^{(0)}\\[5pt]
		&\hspace{2.5cm}+ \Big(- \frac{1}{12\epsilon} + \frac{1}{2}\sigma_4\Big)\eval{Q_2}^{(0)}\,,\\[5pt]
		&\frac{4\pi}{\alpha_s}\eval{Q_2}^{(1)}_{\Sigma;S} = \Big(\frac{20}{\epsilon} - \frac{136}{3} + 8(\sigma_1 - \sigma_3)\Big)\eval{Q_1}^{(0)}\\[5pt]
		&\hspace{2.5cm}+ \Big(\frac{1}{\epsilon}\Big\{-\frac{17}{3} + \frac{8}{3}\xi\Big\} + \frac{62}{9} - \frac{20}{9}\xi - 8(\sigma_2 + \sigma_4)\Big)\eval{Q_2}^{(0)}\,,\\[5pt]
		&\frac{4\pi}{\alpha_s}\eval{\T{Q}_1}^{(1)}_{\T{\Sigma};\T{S}} = \Big(\frac{1}{\epsilon}\Big\{\frac{14}{3} + \frac{8}{3}\xi\Big\} - \frac{76}{9} - \frac{20}{9}\xi
		- \frac{1}{6}\T{\sigma}_1 + \frac{1}{2}\T{\sigma}_4\Big)\eval{\T{Q}_1}^{(0)}\\[5pt]
		&\hspace{2.5cm}+ \Big(- \frac{2}{3\epsilon} - \frac{1}{6}\T{\sigma}_2 + \frac{1}{2}\T{\sigma}_3\Big)\eval{\T{Q}_2}^{(0)}\,,\\[5pt]
		&\frac{4\pi}{\alpha_s}\eval{\T{Q}_2}^{(1)}_{\T{\Sigma};\T{S}} = \Big(-\frac{8}{3\epsilon} - \frac{7}{2} + \frac{1}{4}\T{\sigma}_1 + \frac{7}{12}\T{\sigma}_4\Big)\eval{\T{Q}_1}^{(0)}\\[5pt]
		&\hspace{2.5cm}+ \Big(\frac{1}{\epsilon}\Big\{-\frac{16}{3} + \frac{8}{3}\xi\Big\} + \frac{37}{18} - \frac{20}{9}\xi 
		+ \frac{1}{4}\T{\sigma}_2 + \frac{7}{12}\T{\sigma}_3\Big)\eval{\T{Q}_2}^{(0)}\,,
	\end{split}
\end{equation}
where we have performed the computation in a general $R_\xi$ gauge. Using
\begin{equation}
	\Sigma = \begin{pmatrix}
		0 &&& 0 \\[5pt]
		\sigma_1 &&& - \sigma_2
	\end{pmatrix}\,,
\end{equation}
we find the shift to the tree-level basis transformation
\begin{equation}\label{eq:Niersteshift}
	\begin{split}
		&\Delta_{\T{Q}_1Q_1} = -\frac{2}{3}\sigma_1 - \frac{1}{2}\sigma_3 - \frac{1}{6}\T{\sigma}_1 + \frac{1}{12}\T{\sigma}_2 - \frac{1}{4}\T{\sigma}_3 + \frac{1}{2}\T{\sigma}_4\,,\\[5pt]
		&\Delta_{\T{Q}_1Q_2} = \frac{2}{3}\sigma_2 - \frac{1}{2}\sigma_4 - \frac{1}{48}\T{\sigma}_2 + \frac{1}{16}\T{\sigma}_3 \,,\\[5pt]
		&\Delta_{\T{Q}_2Q_1} = -\frac{37}{12} - \frac{34}{3}\sigma_1 + 20\sigma_2 + \frac{5}{4}\sigma_3 
			+ \frac{1}{4}\T{\sigma}_1 - \frac{1}{8}\T{\sigma}_2 - \frac{7}{24}\T{\sigma}_3 + \frac{7}{12}\T{\sigma}_4\,,\\[5pt]
		&\Delta_{\T{Q}_2Q_2} = -\frac{29}{48} + \frac{1}{12}\sigma_1 + \frac{2}{3}\sigma_2 + \frac{5}{4}\sigma_4 + \frac{1}{32}\T{\sigma}_2 + \frac{7}{96}\T{\sigma}_3\,.
	\end{split}
\end{equation}
Relating Eq.s~\eqref{eq:NiersteCT}~and~\eqref{eq:Niersteshift}, we find
\begin{equation}
	Z_{\T{Q}\T{Q}}^{(1;0)}\,R = - \Delta\,.
\end{equation}

\section{Conclusions}\label{sec:concl}
In this article we have further showcased the utility of loop-corrected Fierz relations by providing a simple method to compute scheme changes using one-loop Fierz identities. Furthermore, we have proven the factorization of renormalization scheme in one-loop Fierz transformations. As a consequence, one may simultaneously perform a change of basis and renormalization scheme by only considering the one-loop corrected transformation of physical operators; the relation between evanescent operator bases is automatically taken into account by the one-loop shifts. In other words, when using one-loop Fierz identities to relate physical operator bases, one also receives the transformation between renormalization schemes for free.

We have illustrated the scheme-factorization in a concrete example, namely the two-loop QCD anomalous dimension matrix relevant for pseudoscalar Higgs coupling contributions to the EDM of the electron. This computation has first been performed in the Larin scheme in \cite{Brod:2023wsh}. In this article we calculated the same ADM, but in the Fierz-conjugated basis. This basis choice allowed the use of the much simpler NDR scheme, where no problematic traces involving $\gamma_5$ occur. After performing the two-loop computation, one-loop Fierz transformations were taken into account to change our results into those for the basis used in \cite{Brod:2023wsh}, transforming from the NDR to the Larin scheme at the same time. The considered example shows the advantage of our method, as it allows to perform the computation in a much simpler scheme, and then change to the target scheme by means of one-loop shifts without worry of accidentally mixing schemes. 

In principle, the traditional, finite-counterterm method can also simultaneously transform between physical bases and renormalization schemes, but at the price of the need to relate all relevant
physical \textit{and evanescent} operators between the two bases, which can be quite non-trivial, as well as the computation of one-loop matrix elements of evanescent operators. In the method
presented in this paper, there is no need to relate evanescent operators between the two bases, and only one-loop matrix elements of physical operators are considered. Although evanescent operators
are technically required in each basis separately to fully project one-loop matrix elements back onto the physical basis, their effect is equivalent to simply introducing a $d$-dimensional analog
of four-dimensional relations. If the one-loop matrix elements of physical operators project onto the tree-level matrix element of an object, $\mathcal{E}$, which does not reduce in $d$-dimensions, but
is given by
\begin{equation}
	\mathcal{E} \overset{d=4}{=} RQ\,,
\end{equation}
then the effect of the projection onto the evanescent operator
\begin{equation}
	E = \mathcal{E} - \big(R + \epsilon \Sigma\big)Q\,,
\end{equation}
in the one-loop basis transformation is equivalent in the respective scheme to the replacement
\begin{equation}
	\eval{\mathcal{E}}^{(0)} = \big(R + \epsilon\Sigma\big)\eval{Q}^{(0)}\,.
\end{equation}

Finally, this simple way of computing higher-order loop corrections is also very well suited to be implemented in computer codes. Hence, the authors plan to include the obtained results in common codes dealing with different operator bases such as \texttt{wilson} \cite{Aebischer:2018bkb} \texttt{WCxf} \cite{Aebischer:2017ugx} and \texttt{abc\_eft} \cite{Proceedings:2019rnh}. Finally, we remark that this method of treating higher-order loop corrections is not limited to the case of Fierz identities, but is in principle applicable to all four-dimensional relations which are altered when continuing to $d\neq 4$.


\section*{Acknowledgements}
\addcontentsline{toc}{section}{\numberline{}Acknowledgements}

We thank Andrzej Buras and Gino Isidori for discussions and useful comments on the manuscript. J.\ A.\,, M.\ P.\ and Z.\ P.\ acknowledge  financial  support  from  the  European  Research  Council  (ERC)  under the European Union's Horizon 2020 research and innovation programme under grant agreement 833280 (FLAY), and from the Swiss National Science Foundation (SNF) under contract 200020-204428.


\clearpage

\appendix


\section{Greek identities}\label{app:greek}

In order to simplify the calculation, the two-loop computation was performed in the $\gamma_5$-basis, containing the operators $\{S_{51},S_{15},V_{51},V_{15},T_{51},T_{15}\}$, which are defined as

\begin{align}\label{eq:S15}
S_{51} =& \big(\bar{b}i\gamma_5 e\big)\big(\bar{e} b\big) \,, &
S_{15} &= \big(\bar{b} e\big)\big(\bar{e}i\gamma_5 b\big)\,, \\
V_{51} =& \big(\bar{b}i\gamma^\mu \gamma_5 e\big)\big(\bar{e}\gamma_\mu b\big)\,,  &
V_{15} &= \big(\bar{b}\gamma^\mu e\big)\big(\bar{e}i\gamma_\mu \gamma_5 b\big)\,, \\[0.5em]
T_{51} =& \big(\bar{b}i\sigma_{\mu \nu}\gamma_5 e\big)\big(\bar{e}\sigma^{\mu \nu} b\big)\,,  &
T_{15} &= \big(\bar{b}\sigma_{\mu \nu} e\big)\big(\bar{e}i\sigma^{\mu \nu}\gamma_5 b\big)\,.\label{eq:T15}
\end{align}
The two operators in \eqref{eq:T15} are not independent in four-dimensions. However, they can differ by evanescent terms in $d$-dimensions. Additionally, using both operators in the basis
simplifies the conversion to the basis Eq.~\eqref{eq:PHYSbasis}.

When inserting these operators into two-loop diagrams we encounter the following structures, which are reduced using standard $d$-dimensional Dirac algebra:

\begin{align}
 \gamma_{\mu}\gamma_{\nu}\gamma_5\otimes \gamma^{\mu}\gamma^{\nu} &= (4-2\epsilon)\gamma_5\otimes \1 -\sigma_{\mu\nu}\gamma_5\otimes\sigma^{\mu\nu} \,,\\
  \gamma_{\mu}\gamma_{\nu}\gamma_5\otimes \gamma^{\nu}\gamma^{\mu} &= (4-2\epsilon)\gamma_5\otimes \1 +\sigma_{\mu\nu}\gamma_5\otimes\sigma^{\mu\nu}\,.
\end{align}

For expressions involving at least three gamma matrices we use the greek reduction \cite{Tracas:1982gp,Buras:1989xd} to reduce them to the following expressions:

\begin{align}
	&\gamma_{\mu}\gamma_{\nu}\gamma_{\alpha}\gamma_{\beta}\gamma_5\otimes\gamma^{\mu}\gamma^{\nu}\gamma^{\alpha}\gamma^{\beta} = (24-112A_1\epsilon)\1\otimes\gamma_5+(40+16A_1\epsilon)\gamma_5\otimes \1 \\[5pt]\notag
	&\hspace{4cm}+(-8+4A_2\epsilon)\sigma_{\mu\nu}\otimes\sigma^{\mu\nu}\gamma_5+(-8+4A_2\epsilon)\sigma_{\mu\nu}\gamma_5\otimes\sigma^{\mu\nu}\,, \\[5pt]
	&\gamma_{\mu}\gamma_{\nu}\gamma_{\alpha}\gamma_5\otimes \gamma^{\mu}\gamma^{\nu}\gamma^{\alpha} = (6 - 4B_1\epsilon)\gamma_\mu\otimes \gamma^\mu\gamma_5 + 10 \gamma_\mu\gamma_5\otimes \gamma^\mu\,,\\[5pt]
	&\gamma_\mu\gamma_\nu\gamma_\rho\gamma_\sigma\gamma_\alpha\gamma_\beta\gamma_5\otimes\gamma^\mu\gamma^\nu\gamma^\rho\gamma^\sigma\gamma^\alpha\gamma^\beta =
	(480 - C_12912\epsilon)\1\otimes\gamma_5+  (544 - C_132\epsilon)\gamma_5\otimes\1 \notag\\[5pt]
	&\hspace{2.5cm}+ (-128 + C_2176\epsilon)\sigma_{\mu\nu}\otimes\sigma^{\mu\nu}\gamma_5 + (-128 + C_2176\epsilon)\sigma_{\mu\nu}\gamma_5\otimes\sigma^{\mu\nu}\,,\\[5pt]
	&\gamma_\mu\gamma_\nu\gamma_\rho\gamma_\sigma\gamma_\alpha\gamma_5\otimes\gamma^\mu\gamma^\nu\gamma^\rho\gamma^\sigma\gamma^\alpha =
(120 - 176D_1\epsilon)\gamma_\mu\otimes\gamma^\mu\gamma_5 + (136 - 48D_1\epsilon)\gamma_\mu\gamma_5\otimes\gamma^\mu \,,
\end{align}
where the arbitrary constants $A_i,B_i,C_i,D_i$ define the renormalization scheme. They are related to the constants $a_s,a_v,a_t$ in the definition of the evanescent operators as follows:

\begin{equation}\label{eq:ais}
	a_s = -96 A_1\,,\quad a_v = 4 B_1 \quad a_t = -8 A_2\,.
\end{equation}

The constants $C_1,C_2,D_1$ must cancel out independently in the calculation of the two-loop ADMs, as is further explained in App.~\ref{app:ren}.

\section{Renormalization}\label{app:ren}

The computation of the two-loop ADM in the NDR basis has been performed in the $\gamma_5$ basis defined in Eqs.~\eqref{eq:S15}-\eqref{eq:T15}. The following renormalization procedure has been employed:
\begin{align}
		\psi^{(0)} &= (1 + \delta Z_{\psi})^{1/2}\,\psi, & {A^{(0)}}^\mu &= (1 + \delta Z_A)^{1/2}A^\mu, \\[0.5em]
		m^{(0)}_{f} &= (1 + \delta Z_{m_{f}})\,m_{f}, & g^{(0)} &= \mu^{\epsilon}(1 + \delta Z_g)\,g\,, \\[0.5em]
		C^{(0)}_i &= C_j\,(\delta_{ji} + \delta Z_{ji})\,, & (m^{(0)})^2 &= \delta Z_{m^2}m^2\,,
\end{align}
where $m$ denotes the IR regulator mass, further discussed in~\cite{Chetyrkin:1997fm}. All of the introduced renormalization constants admit an expansion in the loop-order ($n$) and inverse power of $\epsilon$ ($m$):
\begin{equation}
	\delta Z_i = \sum_{n,m}\frac{1}{\epsilon^m}\delta Z_i^{(n;m)}\,.
\end{equation}

For the standard model counterterms, one finds in general $\xi$-gauge ($N_c =3$)~\cite{Chetyrkin:1997fm,Buras:1998raa}
\begin{equation}
	\begin{split}
		\delta Z_b^{(1;1)} &= -\frac{\alpha_s}{4\pi}\frac{4}{3}\xi_g\,, \\[0.5em]
		\delta Z_{m_b}^{(1;1)} &= -\frac{\alpha_s}{\pi}\,, \\[0.5em]
		\delta Z_G^{(1;1)} &= \frac{\alpha_s}{4\pi}\Big(\frac{13}{2} - \frac{3}{2} \xi_g - \frac{2n_f}{3}\Big)\,, \\[0.5em]
		\delta Z_{m^2}^{(1;1)} &= -\frac{\alpha_s}{4\pi}\Big(\frac{3}{4} + \frac{9}{4} \xi_g + 2n_f\Big)\,, \\[0.5em]
		\delta Z_{g_s}^{(1;1)} &= \frac{\alpha_s}{4\pi}\Big(\frac{n_f}{3} - \frac{11}{2}\Big)\,,
	\end{split}
\end{equation}
and for the physical-physical operators, ordered as $(S_{51},S_{15},V_{51},V_{15},T_{51},T_{15},O_3)$
\begin{equation}
\setlength\arraycolsep{5pt}
	\begin{split}
		\delta Z_{Q_i Q_j}^{(1;1)} &= \frac{\alpha_s}{4\pi}\begin{pmatrix}
			0 & 0 & 0 & 0 & -\frac{1}{3} & 0 & 0 \\[0.5em]
			0 & 0 & 0 & 0 & 0 & -\frac{1}{3} & 0 \\[0.5em]
			0 & 0 & -2 & 2 & 0 & 0 & 0 \\[0.5em]
			0 & 0 & 2 & -2 & 0 & 0 & 0 \\[0.5em]
			-8 & -8 & 0 & 0 & -\frac{16}{3} & \frac{8}{3} & 0 \\[0.5em]
			-8 & -8 & 0 & 0 & \frac{8}{3} & -\frac{16}{3} & 0 \\[0.5em]
			0 & 0 & 0 & 0 & 0 & 0 & 4
		\end{pmatrix}\,.
	\end{split}
\end{equation}
For the physical to evanescent mixing with evanescent ordering as $E_i = (E_{S_{51}}, E_{S_{15}}, E_{V_{51}}, E_{V_{15}})$
\begin{equation}
\setlength\arraycolsep{5pt}
	\delta Z_{Q_iE_j}^{(1;1)} = \frac{\alpha_s}{4\pi}\begin{pmatrix}
		0 & 0 & 0 & 0 \\[0.5em]
		0 & 0 & 0 & 0 \\[0.5em]
		0 & 0 & \frac{1}{3} & 0 \\[0.5em]
		0 & 0 & 0 & \frac{1}{3} \\[0.5em]
		-\frac{1}{3} & 0 & 0 & 0 \\[0.5em]
		0 & -\frac{1}{3} & 0 & 0 \\[0.5em]
		0 & 0 & 0 & 0
	\end{pmatrix}\,,
\end{equation}
where the evanescent operators are defined as
\begin{equation}
	\begin{split}
		E_{S_{51}} =& (\overline b i \gamma^\mu \gamma^\nu \gamma^\rho \gamma^\sigma \gamma_5 e)
			(\overline e \gamma_\mu \gamma_\nu \gamma_\rho \gamma_\sigma b) \\[0.5em]
			& - (40 + 16 A_1\epsilon)S_{51} - (24 - 112 A_1\epsilon)S_{15}
			+ (8 - 4A_2\epsilon)(T_{51} + T_{15})\,, \\[0.5em]
		E_{S_{15}} =& (\overline b \gamma^\mu \gamma^\nu \gamma^\rho \gamma^\sigma e)
			(\overline e i\gamma_\mu \gamma_\nu \gamma_\rho \gamma_\sigma \gamma_5 b) \\[0.5em]
			& - (40 + 16 A_1\epsilon)S_{15} - (24 - 112 A_1\epsilon)S_{51}
			+ (8 - 4A_2\epsilon)(T_{51} + T_{15})\,, \\[0.5em]
		E_{V_{51}} =& (\overline b i\gamma^\mu \gamma^\nu \gamma^\rho\gamma_5 e)
			(\overline e \gamma_\mu \gamma_\nu \gamma_\rho b) - 10 V_{51} - (6 - 4 B_1\epsilon) V_{15}\,, \\[0.5em]
		E_{V_{15}} =& (\overline b \gamma^\mu \gamma^\nu \gamma^\rho e)\,,
			(\overline e i\gamma_\mu \gamma_\nu \gamma_\rho\gamma_5 b) - 10 V_{15} - (6 - 4 B_1\epsilon) V_{51}\,.
	\end{split}
\end{equation}

To avoid an infinite tower of evanescent operators, we need to subtract off the finite mixing of the evanescent operators into
the physical. This will introduce a non-trivial scheme-dependence into the calculation. For a
parameter, $P$, we will break up the scheme-dependent pieces as
\begin{equation}
	P = P^{\text{S.I.}} + A_1\,P^{A_1} + A_2\,P^{A_2} + B_1\,P^{B_1} + C_1\,P^{C_1} + C_2\,P^{C_2} + D_1\,P^{D_1}\,.
\end{equation}
The non-zero contributions to the evanescent $\to$ physical mixing are then given by
\begin{align}
		\delta Z_{E_i Q_j}^{(1;0),\text{S.I.}} &= \frac{\alpha_s}{4\pi}\begin{pmatrix}
			-16 & \frac{688}{3} & 0 & 0 & -16 & -\frac{112}{3} & 0 \\[0.5em]
			\frac{688}{3} & -16 & 0 & 0 & -\frac{112}{3} & -16 & 0 \\[0.5em]
			0 & 0 & 0 & 32 & 0 & 0 & 0 \\[0.5em]
			0 & 0 & 32 & 0 & 0 & 0 & 0
		\end{pmatrix}, \\[0.5em]
		\delta Z_{E_i Q_j}^{(1;0), A_1} &= \frac{\alpha_s}{4\pi}\begin{pmatrix}
			\frac{704}{3} & \frac{1216}{3} & 0 & 0 & \frac{16}{3} & -\frac{112}{3} & 0 \\[0.5em]
			\frac{1216}{3} & \frac{704}{3} & 0 & 0 & -\frac{112}{3} & \frac{16}{3} & 0 \\[0.5em]
			0 & 0 & 0 & 0 & 0 & 0 & 0 \\[0.5em]
			0 & 0 & 0 & 0 & 0 & 0 & 0
		\end{pmatrix}, \\[0.5em]
		\delta Z_{E_i Q_j}^{(1;0), A_2} &= \frac{\alpha_s}{4\pi}\begin{pmatrix}
			64 & 64 & 0 & 0 & -16 & -16 & 0 \\[0.5em]
			64 & 64 & 0 & 0 & -16 & -16 & 0 \\[0.5em]
			0 & 0 & 0 & 0 & 0 & 0 & 0 \\[0.5em]
			0 & 0 & 0 & 0 & 0 & 0 & 0
		\end{pmatrix}, \\[0.5em]
		\delta Z_{E_i Q_j}^{(1;0), B_1} &= \frac{\alpha_s}{4\pi}\begin{pmatrix}
			0 & 0 & 0 & 0 & 0 & 0 & 0 \\[0.5em]
			0 & 0 & 0 & 0 & 0 & 0 & 0 \\[0.5em]
			0 & 0 & 16 & \frac{80}{3} & 0 & 0 & 0 \\[0.5em]
			0 & 0 & \frac{80}{3} & 16 & 0 & 0 & 0
		\end{pmatrix}, \\[0.5em]
		\delta Z_{E_i Q_j}^{(1;0), C_1} &= \frac{\alpha_s}{4\pi}\begin{pmatrix}
			-\frac{32}{3} & -\frac{2912}{3} & 0 & 0 & 0 & 0 & 0 \\[0.5em]
			-\frac{2912}{3} & -\frac{32}{3} & 0 & 0 & 0 & 0 & 0 \\[0.5em]
			0 & 0 & 0 & 0 & 0 & 0 & 0 \\[0.5em]
			0 & 0 & 0 & 0 & 0 & 0 & 0
		\end{pmatrix}, \\[0.5em]
		\delta Z_{E_i Q_j}^{(1;0), C_2} &= \frac{\alpha_s}{4\pi}\begin{pmatrix}
			0 & 0 & 0 & 0 & \frac{176}{3} & \frac{176}{3} & 0 \\[0.5em]
			0 & 0 & 0 & 0 & \frac{176}{3} & \frac{176}{3} & 0 \\[0.5em]
			0 & 0 & 0 & 0 & 0 & 0 & 0 \\[0.5em]
			0 & 0 & 0 & 0 & 0 & 0 & 0
		\end{pmatrix}, \\[0.5em]
		\delta Z_{E_i Q_j}^{(1;0), D_1} &= \frac{\alpha_s}{4\pi}\begin{pmatrix}
			0 & 0 & 0 & 0 & 0 & 0 & 0 \\[0.5em]
			0 & 0 & 0 & 0 & 0 & 0 & 0 \\[0.5em]
			0 & 0 & -16 & -\frac{176}{3} & 0 & 0 & 0 \\[0.5em]
			0 & 0 & -\frac{176}{3} & -16 & 0 & 0 & 0
		\end{pmatrix}.
\end{align}
One also finds spurious $1/\epsilon$ poles in the evanescent to physical mixing
\begin{equation}
	\delta Z_{E_{S_{51}}T_{51}}^{(1;1)} = -\delta Z_{E_{S_{51}}T_{15}}^{(1;1)} 
	= -\delta Z_{E_{S_{15}}T_{51}}^{(1;1)} = \delta Z_{E_{S_{15}}T_{15}}^{(1;1)} = \frac{\alpha_s}{4\pi}\frac{136}{3}\,.
\end{equation}
However, this is a result of choosing a basis which is not unique in $d=4$ and actually corresponds to a mixing into the evanescent operator
\begin{equation}
	T_- = \frac{1}{2}\big(T_{51} - T_{15}\big)\,,
\end{equation}
and so is not problematic.

As an interesting check, we can see that physical operators can only mix into the $E_i$ evanescent operators at one-loop. Therefore,
the shifts can only depend on $A_1$, $A_2$, and $B_1$. In order for the scheme to properly factorize, we immediately know
that all dependence on $C_1$, $C_2$, and $D_1$ must cancel when computing the ADM in this basis. 

For the two-loop QCD calculation, only the $b$-quark and $m_b$ counterterms at two loops are required. These are given by~\cite{Buras:2020xsm,Buras:1998raa,Brod:2020lhd}

\begin{equation}
	\begin{split}
		\delta Z_b^{(2)} =& \Big(\frac{\alpha_s}{4\pi}\Big)^2\Bigg[\frac{1}{\epsilon^2}\Big(3\xi_g + \frac{17}{9}\xi_g^2\Big)
			+ \frac{1}{\epsilon}\Big(-\frac{67}{6} + \frac{2}{3}n_f - 4\xi_g - \frac{1}{2}\xi_g^2\Big)\Bigg]\,,\\[0.5em]
		\delta Z_{m_b}^{(2)} =& \Big(\frac{\alpha_s}{4\pi}\Big)^2\Bigg[\frac{1}{\epsilon^2}\Big(30 - \frac{4}{3}n_f\Big)
			+ \frac{1}{\epsilon}\Big(-\frac{101}{3} + \frac{10}{9}n_f\Big)\Bigg]\,.
	\end{split}
\end{equation}

With this, the (scheme-independent) $1/\epsilon^2$ pole of the operator counterterm is given by (setting $n_f$ = 5)
\begin{equation}
\setlength\arraycolsep{5pt}
		\delta Z^{(2;2)}_{Q_iQ_j} = \Big(\frac{\alpha_s}{4\pi}\Big)^2\begin{pmatrix}
			\frac{4}{3} & \frac{4}{3} & 0 & 0 & \frac{13}{6} & -\frac{4}{9} & 0 \\[5pt]
			\frac{4}{3} & \frac{4}{3} & 0 & 0 & -\frac{4}{9} & \frac{13}{6} & 0 \\[5pt]
			0 & 0 & \frac{35}{3} & - \frac{35}{3} & 0 & 0 & 0 \\[5pt]
			0 & 0 & -\frac{35}{3} & \frac{35}{3} & 0 & 0 & 0 \\[5pt]
			\frac{124}{3} & \frac{124}{3} & 0 & 0 & 32 & -\frac{140}{9} & 0 \\[5pt]
			\frac{124}{3} & \frac{124}{3} & 0 & 0 & -\frac{140}{9} & 32 & 0 \\[5pt]
			0 & 0 & 0 & 0 & 0 & 0 & - \frac{22}{3}
	\end{pmatrix}\,.
\end{equation}
The $1/\epsilon$ poles are scheme-dependent in general and are given by
\begin{equation}
	\setlength\arraycolsep{5pt}
		\delta Z^{(2;1),\text{S.I.}}_{Q_i Q_j} = \Big(\frac{\alpha_s}{4\pi}\Big)^2\begin{pmatrix}
			\frac{65}{9} & \frac{2}{3} & -\frac{1}{2} & -\frac{1}{2} & -\frac{287}{108} & \frac{1}{9} & 0 \\[5pt]
			\frac{2}{3} & \frac{65}{9} & \frac{1}{2} & \frac{1}{2} & \frac{1}{9} & -\frac{287}{108} & 0 \\[5pt]
			-2 & 2 & -\frac{271}{18} & \frac{239}{18} & 0 & 0 & 0 \\[5pt]
			-2 & 2 & \frac{239}{18} & -\frac{271}{18} & 0 & 0 & 0 \\[5pt]
			\frac{266}{9} & -\frac{478}{9} & 0 & 0 & -\frac{1085}{27} & \frac{490}{27} & 0 \\[5pt]
			-\frac{478}{9} & \frac{266}{9} & 0 & 0 & \frac{490}{27} & -\frac{1085}{27} & 0 \\[5pt]
			0 & 0 & 0 & 0 & 0 & 0 & \frac{253}{9}
		\end{pmatrix}\,,
\end{equation}
\begin{equation}
	\setlength\arraycolsep{5pt}
		\delta Z^{(2;1),A_1}_{Q_i Q_j} = \Big(\frac{\alpha_s}{4\pi}\Big)^2\begin{pmatrix}
			\frac{8}{9} & -\frac{56}{9} & 0 & 0 & 0 & 0 & 0 \\[5pt]
			-\frac{56}{9} & \frac{8}{9} & 0 & 0 & 0 & 0 & 0 \\[5pt]
			0 & 0 & 0 & 0 & 0 & 0 & 0 \\[5pt]
			0 & 0 & 0 & 0 & 0 & 0 & 0 \\[5pt]
			\frac{136}{3} & -\frac{952}{3} & 0 & 0 & -\frac{16}{9} & \frac{112}{9} & 0 \\[5pt]
			-\frac{952}{3} & \frac{136}{3} & 0 & 0 & \frac{112}{9} & -\frac{16}{9} & 0 \\[5pt]
			0 & 0 & 0 & 0 & 0 & 0 & 0
		\end{pmatrix}\,,
\end{equation}
\begin{equation}
	\setlength\arraycolsep{5pt}
		\delta Z^{(2;1),A_2}_{Q_i Q_j} = \Big(\frac{\alpha_s}{4\pi}\Big)^2\begin{pmatrix}
			0 & 0 & 0 & 0 & \frac{2}{9} & \frac{2}{9} & 0 \\[5pt]
			0 & 0 & 0 & 0 & \frac{2}{9} & \frac{2}{9} & 0 \\[5pt]
			0 & 0 & 0 & 0 & 0 & 0 & 0 \\[5pt]
			0 & 0 & 0 & 0 & 0 & 0 & 0 \\[5pt]
			-\frac{64}{3} & -\frac{64}{3} & 0 & 0 & \frac{70}{9} & \frac{70}{9} & 0 \\[5pt]
			-\frac{64}{3} & -\frac{64}{3} & 0 & 0 & \frac{70}{9} & \frac{70}{9} & 0 \\[5pt]
			0 & 0 & 0 & 0 & 0 & 0 & 0
		\end{pmatrix}\,,
\end{equation}
\begin{equation}
	\setlength\arraycolsep{5pt}
		\delta Z^{(2;1),B_1}_{Q_i Q_j} = \Big(\frac{\alpha_s}{4\pi}\Big)^2\begin{pmatrix}
			0 & 0 & 0 & 0 & 0 & 0 & 0 \\[5pt]
			0 & 0 & 0 & 0 & 0 & 0 & 0 \\[5pt]
			0 & 0 & \frac{8}{3} & \frac{86}{9} & 0 & 0 & 0 \\[5pt]
			0 & 0 & \frac{86}{9} & \frac{8}{3} & 0 & 0 & 0 \\[5pt]
			0 & 0 & 0 & 0 & 0 & 0 & 0 \\[5pt]
			0 & 0 & 0 & 0 & 0 & 0 & 0 \\[5pt]
			0 & 0 & 0 & 0 & 0 & 0 & 0
		\end{pmatrix}\,,
\end{equation}
\begin{equation}
	\setlength\arraycolsep{5pt}
		\delta Z^{(2;1),C_1}_{Q_i Q_j} = \Big(\frac{\alpha_s}{4\pi}\Big)^2\begin{pmatrix}
			0 & 0 & 0 & 0 & 0 & 0 & 0 \\[5pt]
			0 & 0 & 0 & 0 & 0 & 0 & 0 \\[5pt]
			0 & 0 & 0 & 0 & 0 & 0 & 0 \\[5pt]
			0 & 0 & 0 & 0 & 0 & 0 & 0 \\[5pt]
			\frac{16}{9} & \frac{1456}{9} & 0 & 0 & 0 & 0 & 0 \\[5pt]
			\frac{1456}{9} & \frac{16}{9} & 0 & 0 & 0 & 0 & 0 \\[5pt]
			0 & 0 & 0 & 0 & 0 & 0 & 0
		\end{pmatrix}\,,
\end{equation}
\begin{equation}
	\setlength\arraycolsep{5pt}
		\delta Z^{(2;1),C_2}_{Q_i Q_j} = \Big(\frac{\alpha_s}{4\pi}\Big)^2\begin{pmatrix}
			0 & 0 & 0 & 0 & 0 & 0 & 0 \\[5pt]
			0 & 0 & 0 & 0 & 0 & 0 & 0 \\[5pt]
			0 & 0 & 0 & 0 & 0 & 0 & 0 \\[5pt]
			0 & 0 & 0 & 0 & 0 & 0 & 0 \\[5pt]
			0 & 0 & 0 & 0 & -\frac{88}{9} & -\frac{88}{9} & 0 \\[5pt]
			0 & 0 & 0 & 0 & -\frac{88}{9} & -\frac{88}{9} & 0 \\[5pt]
			0 & 0 & 0 & 0 & 0 & 0 & 0
		\end{pmatrix}\,,
\end{equation}
\begin{equation}
	\setlength\arraycolsep{5pt}
		\delta Z^{(2;1),D_1}_{Q_i Q_j} = \Big(\frac{\alpha_s}{4\pi}\Big)^2\begin{pmatrix}
			0 & 0 & 0 & 0 & 0 & 0 & 0 \\[5pt]
			0 & 0 & 0 & 0 & 0 & 0 & 0 \\[5pt]
			0 & 0 & -\frac{8}{3} & -\frac{88}{9} & 0 & 0 & 0 \\[5pt]
			0 & 0 & -\frac{88}{9} & -\frac{8}{3} & 0 & 0 & 0 \\[5pt]
			0 & 0 & 0 & 0 & 0 & 0 & 0 \\[5pt]
			0 & 0 & 0 & 0 & 0 & 0 & 0 \\[5pt]
			0 & 0 & 0 & 0 & 0 & 0 & 0
		\end{pmatrix}\,.
\end{equation}

As a next step, we will transform to the physical basis defined in Eq.~\eqref{eq:PHYSbasis}. This reduces the operator basis from seven
operators to just four, but we can artificially enlarge the basis in order to have an invertible transformation
between the bases. The enlarged basis is given by
$\{\tilde{\mathcal{O}}_s, \tilde{\mathcal{O}}_v, \tilde{\mathcal{O}}_t, \tilde{\mathcal{O}}_3, 
S_-, V_+, T_-\}$ where the first four operators are defined in Eq.~\eqref{eq:PHYSbasis} and the artificial operators are given by
\begin{equation}
	S_- = \frac{1}{2}\big(S_{51} - S_{15}\big), \quad V_+ = \frac{1}{2}\big(V_{51} + V_{15}\big), 
	\quad T_- = \frac{1}{2}\big(T_{51} - T_{15}\big)\,.
\end{equation}
As another check, the counterterms should factorize between the physical and artificial operators and there should be no
mixing between the two sectors. Using a similar procedure for the evanescent operators to project onto those introduced in
Eq.~\eqref{eq:EVsNDR} gives the physical to evanescent and evanescent to physical counterterms in the relevant basis.

We then find for the one-loop physical-to-physical counterterms
\begin{equation}
	\setlength\arraycolsep{5pt}
	\delta Z^{(1;1)}_{Q_iQ_j} = \frac{\alpha_s}{4\pi}\begin{pmatrix}
		0 & 0 & -\frac{1}{3} & 0 \\
		0 & -4 & 0 & 0 \\
		-16 & 0 & -\frac{8}{3} & 0 \\
		0 & 0 & 0 & 4
	\end{pmatrix}\,,
\end{equation}
for the one-loop physical-to-evanescent counterterms (using $E_i = (\tilde{E}_t, \tilde{E}_v)$)
\begin{equation}
	\setlength\arraycolsep{5pt}
	\delta Z^{(1;1)}_{Q_iE_j} = \frac{\alpha_s}{4\pi}\begin{pmatrix}
		0 & 0 \\
		0 & \frac{1}{3} \\
		-\frac{1}{3} & 0 \\
		0 & 0
	\end{pmatrix}\,,
\end{equation}
and for the one-loop evanescent-to-physical counterterms
\begin{align}
	\setlength\arraycolsep{5pt}
	\delta Z^{(1;0),\text{S.I.}}_{E_iQ_j} = \frac{\alpha_s}{4\pi}\begin{pmatrix}
		\frac{640}{3} & 0 & -\frac{160}{3} & 0 \\
		0 & -32 & 0 & 0
	\end{pmatrix},&\quad
	\setlength\arraycolsep{5pt}
	\delta Z^{(1;0),A_2}_{E_iQ_j} = \frac{\alpha_s}{4\pi}\begin{pmatrix}
		640 & 0 & -32 & 0 \\
		0 & 0 & 0 & 0
	\end{pmatrix}, \\[10pt]
	\setlength\arraycolsep{5pt}
	\delta Z^{(1;0),A_2}_{E_iQ_j} = \frac{\alpha_s}{4\pi}\begin{pmatrix}
		128 & 0 & -32 & 0 \\
		0 & 0 & 0 & 0
	\end{pmatrix},& \quad
	\setlength\arraycolsep{5pt}
	\delta Z^{(1;0),B_1}_{E_iQ_j} = \frac{\alpha_s}{4\pi}\begin{pmatrix}
		0 & 0 & 0 & 0 \\
		0 & -\frac{32}{3} & 0 & 0
	\end{pmatrix}, \\[10pt]
	\setlength\arraycolsep{5pt}
	\delta Z^{(1;0),C_1}_{E_iQ_j} = \frac{\alpha_s}{4\pi}\begin{pmatrix}
		-\frac{2944}{3} & 0 & 0 & 0 \\
		0 & 0 & 0 & 0
	\end{pmatrix},& \quad
	\setlength\arraycolsep{5pt}
	\delta Z^{(1;0),C_2}_{E_iQ_j} = \frac{\alpha_s}{4\pi}\begin{pmatrix}
		0 & 0 & \frac{352}{3} & 0 \\
		0 & 0 & 0 & 0
	\end{pmatrix}, \\[10pt]
	\setlength\arraycolsep{5pt}
	\delta Z^{(1;0),D_2}_{E_iQ_j} = \frac{\alpha_s}{4\pi}&\begin{pmatrix}
		0 & 0 & 0 & 0 \\
		0 & \frac{128}{3} & 0 & 0
	\end{pmatrix}\,.
\end{align}
For the two-loop counterterms, one finds
\begin{equation}
	\setlength\arraycolsep{5pt}
	\delta Z^{(2;2)}_{Q_iQ_j} = \Big(\frac{\alpha_s}{4\pi}\Big)^2\begin{pmatrix}
		\frac{8}{3} & 0 & \frac{31}{18} & 0 \\
		0 & \frac{70}{3} & 0 & 0 \\
		\frac{248}{3} & 0 & \frac{148}{9} & 0 \\
		0 & 0 & 0 & -\frac{22}{3}
	\end{pmatrix}\,,
\end{equation}
for the $1/\epsilon^2$ poles and
\begin{align}
	\setlength\arraycolsep{5pt}
	\delta Z^{(2;1),\text{S.I.}}_{Q_iQ_j} &= \Big(\frac{\alpha_s}{4\pi}\Big)^2\begin{pmatrix}
		\frac{71}{9} & 0 & -\frac{275}{108} & 0 \\
		0 & -\frac{85}{3} & 0 & 0 \\
		-\frac{212}{9} & 0 & -\frac{595}{27} & 0 \\
		0 & 0 & 0 & \frac{253}{9}
	\end{pmatrix},\\[10pt]
	\setlength\arraycolsep{5pt}
	\delta Z^{(2;1),A_1}_{Q_iQ_j} &= \Big(\frac{\alpha_s}{4\pi}\Big)^2\begin{pmatrix}
		-\frac{16}{3} & 0 & 0 & 0 \\
		0 & 0 & 0 & 0 \\
		-272 & 0 & \frac{32}{3} & 0 \\
		0 & 0 & 0 & 0
	\end{pmatrix},\\[10pt]
	\setlength\arraycolsep{5pt}
	\delta Z^{(2;1),A_2}_{Q_iQ_j} &= \Big(\frac{\alpha_s}{4\pi}\Big)^2\begin{pmatrix}
		0 & 0 & \frac{4}{9} & 0 \\
		0 & 0 & 0 & 0 \\
		-\frac{128}{3} & 0 & \frac{140}{9} & 0 \\
		0 & 0 & 0 & 0
	\end{pmatrix},\\[10pt]
	\setlength\arraycolsep{5pt}
	\delta Z^{(2;1),B_1}_{Q_iQ_j} &= \Big(\frac{\alpha_s}{4\pi}\Big)^2\begin{pmatrix}
		0 & 0 & 0 & 0 \\
		0 & -\frac{62}{9} & 0 & 0 \\
		0 & 0 & 0 & 0 \\
		0 & 0 & 0 & 0
	\end{pmatrix},\\[10pt]
	\setlength\arraycolsep{5pt}
	\delta Z^{(2;1),C_1}_{Q_iQ_j} &= \Big(\frac{\alpha_s}{4\pi}\Big)^2\begin{pmatrix}
		0 & 0 & 0 & 0 \\
		0 & 0 & 0 & 0 \\
		\frac{1472}{9} & 0 & 0 & 0 \\
		0 & 0 & 0 & 0
	\end{pmatrix},\\[10pt]
	\setlength\arraycolsep{5pt}
	\delta Z^{(2;1),C_2}_{Q_iQ_j} &= \Big(\frac{\alpha_s}{4\pi}\Big)^2\begin{pmatrix}
		0 & 0 & 0 & 0 \\
		0 & 0 & 0 & 0 \\
		0 & 0 & -\frac{176}{9} & 0 \\
		0 & 0 & 0 & 0
	\end{pmatrix},\\[10pt]
	\setlength\arraycolsep{5pt}
	\delta Z^{(2;1),D_1}_{Q_iQ_j} &= \Big(\frac{\alpha_s}{4\pi}\Big)^2\begin{pmatrix}
		0 & 0 & 0 & 0 \\
		0 & \frac{64}{9} & 0 & 0 \\
		0 & 0 & 0 & 0 \\
		0 & 0 & 0 & 0
	\end{pmatrix},
\end{align}
for the $1/\epsilon$ poles.

Finally, the ADM can be found by requiring that the bare parameters are independent of the arbitrary renormalization scale
\begin{equation}
	\frac{d C_i^{(\text{bare})}}{d\log\mu} = \frac{d C_j}{d\log\mu}\,Z_{ji} + C_j\frac{dZ_{ji}}{d\log\mu} = 0\,.
\end{equation}
Rearranging gives
\begin{equation}
	\frac{d C_i}{d\log\mu} = C_j\gamma_{ji} \quad \Rightarrow \quad \gamma_{ij} = -\frac{dZ_{ik}}{d\log\mu}\,Z^{-1}_{kj}
	= -\beta(\alpha_s,\epsilon)\frac{dZ_{ik}}{d\alpha_s}\,Z^{-1}_{kj}\,.
\end{equation}
In the above, one needs to use the $d$-dimensional beta function
\begin{equation}
	\frac{d\alpha_s}{d\log\mu} = \beta(\alpha_s,\epsilon) = -2\epsilon\alpha_s + \beta(\alpha_s)\,,
\end{equation}
due to the presence of $1/\epsilon$ poles in the counterterm. Otherwise one would end up with unregulated divergences in the ADM. We can now expand
\begin{equation}
	Z^{-1}_{ij} = \delta_{ij} - \frac{\alpha_s}{4\pi}\delta Z^{(1)}_{ij},\quad
	\beta(\alpha_s) = - 2\alpha_s\Big(\frac{\alpha_s}{4\pi}\Big)\beta_0\,,
\end{equation}
where $\beta_0=11-\frac{2}{3}n_f$. Therefore, one finds
\begin{equation}
	\begin{split}
		\gamma^{(0)}_{ij} =& 2\epsilon\,\delta Z^{(1)}_{ij}\,, \\[5pt]
		\gamma^{(1)}_{ij} =& 2\epsilon\,\Big(2\delta Z^{(2)}_{ij} - \delta Z^{(1)}_{ik}\delta Z^{(1)}_{kj}\Big)
			+ 2\beta_0\,\delta Z^{(1)}_{ij}\,.
	\end{split}
\end{equation}
Expanding again in poles, in order for the physical ADM to be finite, the two-loop $1/\epsilon^2$ poles 
and the one-loop $1/\epsilon$ poles must be related via 
\begin{equation}
	2\delta Z^{(2;2)}_{Q_iQ_j} = \delta Z^{(1;1)}_{Q_iQ_k}\delta Z^{(1;1)}_{Q_kQ_j} - \beta_0\delta Z^{(1;1)}_{Q_iQ_j}\,,
\end{equation}
giving the final expressions for the one- and two-loop physical ADM
\begin{equation}
	\begin{split}
		\gamma^{(0)}_{Q_iQ_j} =& 2\delta Z^{(1;1)}_{Q_iQ_j}\,, \\[5pt]
		\gamma^{(1)}_{Q_iQ_j} =& 4\delta Z^{(2;1)}_{Q_iQ_j} - 2\delta Z^{(1;1)}_{Q_iE_k}\delta Z^{(1;0)}_{E_kQ_j}\,.
	\end{split}
\end{equation}
Using our counterterms in these expressions, we find
\begin{equation}
	\setlength\arraycolsep{5pt}
	\gamma^{(0)}_{ij} = \begin{pmatrix}
		0 & 0 & -\frac{2}{3} & 0 \\
		0 & -8 & 0 & 0 \\
		-32 & 0 & -\frac{16}{3} & 0 \\
		0 & 0 & 0 & 8
	\end{pmatrix}\,,
\end{equation}
for the one-loop ADM and
\begin{equation}
	\begin{split}
	\setlength\arraycolsep{5pt}
		\gamma^{(1),\text{S.I.}}_{ij} = \begin{pmatrix}
			\frac{284}{9} & 0 & -\frac{275}{27} & 0 \\
			0 & -92 & 0 & 0 \\
			48 & 0 & -\frac{3340}{27} & 0 \\
			0 & 0 & 0 & \frac{1012}{9}
		\end{pmatrix}\,,& \quad
	\setlength\arraycolsep{5pt}
		\gamma^{(1),A_1}_{ij} =\begin{pmatrix}
			-\frac{64}{3} & 0 & 0 & 0 \\
			0 & 0 & 0 & 0 \\
			-\frac{1984}{3} & 0 & \frac{64}{3} & 0 \\
			0 & 0 & 0 & 0
		\end{pmatrix}\,, \\[10pt]
	\setlength\arraycolsep{5pt}
		\gamma^{(1),A_2}_{ij} = \begin{pmatrix}
			0 & 0 & \frac{16}{9} & 0 \\
			0 & 0 & 0 & 0 \\
			-\frac{256}{3} & 0 & \frac{368}{9} & 0 \\
			0 & 0 & 0 & 0
		\end{pmatrix}\,,& \quad
	\setlength\arraycolsep{5pt}
		\gamma^{(1),B_1}_{ij} =\begin{pmatrix}
			0 & 0 & 0 & 0 \\
			0 & -\frac{184}{9} & 0 & 0 \\
			0 & 0 & 0 & 0 \\
			0 & 0 & 0 & 0
		\end{pmatrix}\,,
	\end{split}
\end{equation}
for the two-loop ADMs. As expected, all dependence on the scheme variables $C_1$, $C_2$, and $D_1$ has dropped out of the
ADMs, as it must be for the scheme to factorize. Also note that this is exactly the same as the ADM found previously
when using the relations in Eq.~\eqref{eq:ais}. Using
the above matrices along with the corresponding shifts, one again reproduces the ADMs found in Ref.~\cite{Brod:2023wsh}.

\bibliographystyle{JHEP}

\bibliography{refs}

\end{document}